\documentclass[10pt,journal,compsoc,twoside]{IEEEtran}

\newcommand{\ignore}[1]{}
\ifCLASSOPTIONcompsoc
  \usepackage[nocompress]{cite}
\else
  \usepackage{cite}
\fi
\usepackage{algorithm}
\usepackage{algorithmic}
\usepackage{makecell}
\usepackage{bm}
\usepackage{graphicx}
\usepackage{multirow}
\usepackage{makecell}
\usepackage{amssymb}
\usepackage{threeparttable}
\usepackage{bm}
\usepackage{xcolor}
\usepackage{url}
\usepackage{color}
\usepackage{subfig}

\ifCLASSINFOpdf
\else
\fi
\begin{document}
%
\title{Detecting Adversarial Image Examples in Deep Neural Networks with Adaptive Noise Reduction}
%
%
%
%

\author{Bin Liang, Hongcheng Li, Miaoqiang Su, Xirong Li, Wenchang Shi and Xiaofeng Wang

\IEEEcompsocitemizethanks{
\IEEEcompsocthanksitem B. Liang, H. Li, M. Su, X. Li and W. Shi are with the School of Information, Renmin University of China, Beijing, China; and also with Key Laboratory of DEKE (Renmin University of China), MOE, China. Email: \{liangb, owenlee, sumiaoqiang, xirong, wenchang\}@ruc.edu.cn.

\IEEEcompsocthanksitem X. Wang is with the Center for Security Informatics, Indiana University Bloomington, Bloomington, IN 47405 USA. Email: xw7@indiana.edu.
}
}
%
%

\markboth{IEEE TRANSACTIONS ON DEPENDABLE AND SECURE COMPUTING, MANUSCRIPT ID}%
{Liang \MakeLowercase{\textit{et al.}}: Detecting Adversarial Image Examples in Deep Neural Networks with Adaptive Noise Reduction}
%



\IEEEtitleabstractindextext{%
\begin{abstract}
Recently, many studies have demonstrated deep neural network (DNN) classifiers can be fooled by the adversarial example, which is crafted via introducing some perturbations into an original sample. Accordingly, some powerful defense techniques were proposed. However, existing defense techniques often require modifying the target model or depend on the prior knowledge of attacks. In this paper, we propose a straightforward method for detecting adversarial image examples, which can be directly deployed into unmodified off-the-shelf DNN models. We consider the perturbation to images as a kind of noise and introduce two classic image processing techniques, \textit{scalar quantization} and \textit{smoothing spatial filter}, to reduce its effect. The image entropy is employed as a metric to implement an adaptive noise reduction for different kinds of images. Consequently, the adversarial example can be effectively detected by comparing the classification results of a given sample and its denoised version, without referring to any prior knowledge of attacks. More than 20,000 adversarial examples against some state-of-the-art DNN models are used to evaluate the proposed method, which are crafted with different attack techniques. The experiments show that our detection method can achieve a high overall F1 score of 96.39\% and certainly raises the bar for defense-aware attacks.
\end{abstract}

}

\maketitle

\IEEEdisplaynontitleabstractindextext

%
\IEEEpeerreviewmaketitle

\IEEEraisesectionheading{\section{Introduction}\label{sec:introduction}}

%
%
%
%
\IEEEPARstart{D}{eep} neural networks (DNNs) \cite{liu2017survey} have been widely adopted in many applications such as computer vision~\cite{lecun2010convolutional,szegedy2015going}, speech recognition~\cite{dahl2012context,hinton2012deep}, and natural language processing~\cite{collobert2008unified,zhang2015character}. DNNs have exhibited very impressive performance in these tasks, especially in the image classification~\cite{szegedy2015going}. Some DNN-based classifiers achieved even better performance than human~\cite{sun2014deep,sun2014deep2}. Meanwhile, their robustness has also raised concerns.

Some recent studies~\cite{goodfellow2014explaining,moosavi2016deepfool,szegedy2013intriguing} demonstrate that DNN-based image classifiers can be fooled by \textit{adversarial examples}, which are well-crafted to cause a trained model to misclassify. For example, as presented in \cite{goodfellow2014explaining}, added with some imperceptible perturbations, an image can be misclassified with very high confidence by GoogLeNet~\cite{szegedy2015going}, while a human observer can still correctly classify it without noticing the existence of the introduced perturbations. These studies indicate that the adversaries could potentially use the crafted image to inflict serious damages.  As shown in~\cite{papernot2016practical}, a \textit{stop} sign, after being crafted, will be incorrectly classified as a \textit{yield} sign. As a result, a self-driving car equipped with the DNN classifier may behave dangerously.

Some techniques have been proposed to defend adversarial examples in DNNs~\cite{goodfellow2014explaining,kereliuk2015deep,papernot2016limitations,papernot2016distillation}. Most of them require modifying the target model. For example, the \textit{adversarial training} is a straightforward defense technique which uses as many adversarial examples as possible during training process as a kind of regularization~\cite{goodfellow2014explaining,kereliuk2015deep,papernot2016limitations}. Papernot et al. ~\cite{papernot2016distillation} introduced a defense technique named \textit{defensive distillation}. Two networks were trained as a distillation, where the first network produced probability vectors to label the original dataset, while the other was trained using the newly labeled dataset. As a result, the effectiveness of adversarial examples can be substantially reduced. Several very recent studies~\cite{feinman2017detecting,metzen2017detecting,grosse2017statistical,xu2017feature} focus on detecting adversarial examples directly. Similarly, these techniques also require modifying the model or acquiring sufficient adversarial examples, such as training new sub-models~\cite{feinman2017detecting}, retraining a revised model as a detector using known adversarial examples~\cite{metzen2017detecting}, performing a statistical test on a large group of adversarial and benign examples~\cite{grosse2017statistical}, or training the key detection parameter using a number of adversarial examples and their corresponding benign ones~\cite{xu2017feature}.

Unfortunately, retraining an existing model or changing its architecture will introduce expensive training cost. Generating appropriate adversarial examples for training or statistical testing is also of high cost and depends on a very comprehensive prior knowledge of various potential adversarial techniques. Even worse, the attacker can craft adversarial examples with the  technique unknown to the defender. In this case, the adversarial example has a good chance to evade the classifier. Moreover, training a classifier with an emerging attack technique would take some time. There always is a window for attackers to craft effectual adversarial examples. Furthermore, most of existing defense techniques are model-specific. To apply a defense technique to different models, they need to be rebuilt or retrained individually. The security enhancement to a model cannot be directly applied to other ones.\footnote{For the latest version of this paper, please referring to \textcolor{blue}{\url{https://ieeexplore.ieee.org/document/8482346}}.}

To address the aforementioned challenges, we present in this paper a new technique capable of effectively capturing adversarial examples, even in the absence of prior knowledge about potential attacks.

Our approach is based upon the observation that to make the adversarial change imperceptible, the perturbation incurred by the adversarial examples typically needs to be confined within a small range.  This is important, since otherwise, the example will be easily identified by human. Consequently, the information introduced by the perturbation should also be less than that of the original image. In the proposed method, we treat the perturbation as a kind of artificial noise and leverage the noise reduction techniques to reduce its adversarial effect. If the effect is downgraded properly, the denoised adversarial example will be classified as a new class that is different from the adversarial target. On the other hand, for the legitimate sample, the same denoising operation will most likely just slightly changes the  image's semantics, keeping it still within its original category. Intuitively, all the adversarial perturbation is added later on to the image and therefore tends to less tolerant of the noise reduction process than the original image information. The information remaining in a denoised benign sample can be still enough for the classifier to correctly identify its class. In fact, some studies ~\cite{goodfellow2009measuring,lecun1989backpropagation}  have shown that the state-of-the-art classifiers are somewhat robust against certain degree of distortions. To this end, the adversarial example can be effectively detected by inspecting whether the classification of a sample is changed after it is denoised.

\begin{figure}[t]
\centering
\includegraphics{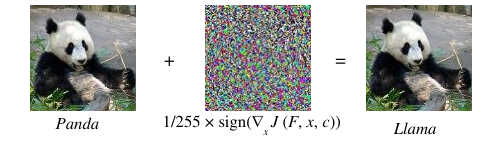}
\caption{Very small perturbation can result in an effective adversarial example for a color image.}
\label{Eps_ImageNet}
\end{figure}

\begin{figure}[t]
\centering
\includegraphics{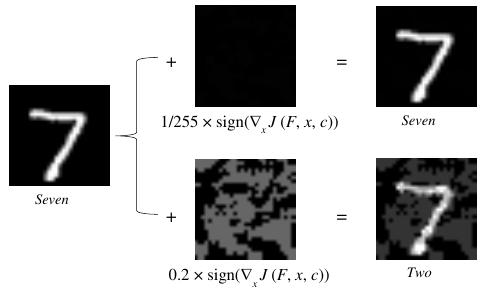}
\caption{Only large perturbation can produce an adversarial example for a simple image.}
\label{Eps_MNIST}
\end{figure}

Two classic image processing techniques, \textit{scalar quantization} and \textit{smoothing spatial filter}, are leveraged to reduce the effect of perturbations. However, it is actually inappropriate to denoise all samples in the same way. A simple image, e.g., a grayscale handwritten digit, may only consist of hundreds of pixels. In other words, the perturbation space is limited when generating an adversarial example for this kind of image. The adversary have to magnify the perturbation to generate an effective adversarial example \cite{goodfellow2014explaining}. On the other hand, a color photo can often provide a larger pertubation space. Adding a small scale pertubation can make it be misclassified. As Fig. \ref{Eps_ImageNet} illustrates, manipulating a color image with very small perturbation can  generate an effective adversarial example with the attack method in \cite{goodfellow2014explaining}. However, using the same strength perturbation fails to generate adversarial example for a handwriting digit, as Fig. \ref{Eps_MNIST} shows. Only when we magnify the perturbation to at least 50 times, can we get an effective adversarial example.
As a result, the quantization or smoothing suitable for a high-resolution sample may not be enough for a low-resolution one. Using a uniform noise reduction strategy may over-denoise some samples or under-denoise some others, and introduce false positives and negatives. Obviously, we should employ different denoising strategies for different images.

\begin{figure}[tb]
\centering
\includegraphics{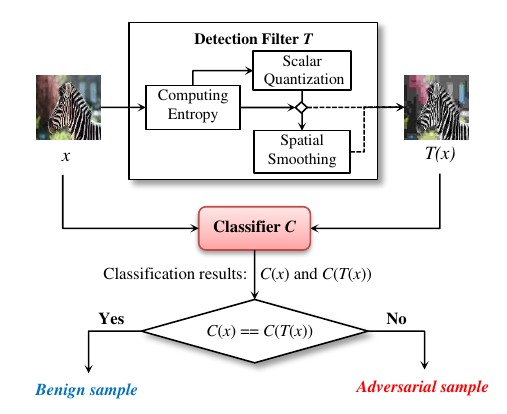}
\caption{Detection Framework.}
\label{Fig2}
\end{figure}

To improve the generality of our method for different kinds of images, an adaptive noise reduction is enforced. The \textit{entropy} of the image is utilized to measure the complexity of samples. An aggressive noise reduction strategy is adopted for the sample with a low entropy, which may be perturbed heavily by the adversary. For high-entropy samples, a light strategy is employed. Specifically, as illustrated in Fig. \ref{Fig2}, the key component of our method is a detection filter \textit{T}. When feeding a sample \textit{x} to the target classifier \textit{C}, it will be denoised by \textit{T} to generate a filtered sample \textit{T}(\textit{x}). The sample
is first quantized with an appropriate interval size, which is determined by computing the entropy of the sample. We also use the entropy to decide whether the quantized sample needs to be smoothed. Only when the entropy is larger than
a threshold, will it be smoothed by a spatial smoothing filter. Finally, two individual classifications are performed. The one is for the original sample \textit{x}, and the other is for the denoised sample \textit{T}(\textit{x}). Only when the two classifications are equal, i.e., \textit{C}(\textit{x}) == \textit{C}(\textit{T}(\textit{x})), will \textit{x} be considered benign. Otherwise, it is identified adversarial. In practice, the detection mechanism cannot know the true class of \textit{x} in advance to validate its  current classification \textit{C}(\textit{x}). In our method, \textit{C}(\textit{T}(\textit{x})) actually acts as a detection baseline to detect potential adversarial examples.

We employ some state-of-the-art DNN models and popular datasets, such as GoogLeNet~\cite{szegedy2015going}, CaffeNet~\cite{jia2014caffe}, MNIST~\cite{URL:TheMNISTDatabaseOfHandwrittenDigits} and ImageNet~\cite{deng2009imagenet} to evaluate the effectiveness of the proposed method. Three up-to-date attack techniques, i.e., FGSM~\cite{goodfellow2014explaining}, DeepFool~\cite{moosavi2016deepfool}, and CW attacks~\cite{carlini2016towards}, are used to craft adversarial examples. We evaluate our method in two attack scenarios, namely, \textit{defense-unaware} and \textit{defense-aware}. In the defense-unaware scenario, we generated 21,673 effectual adversarial examples in total. The experiments show that the proposed method can achieve an overall recall of 95.00\% and an overall precision of 97.81\% for detecting the adversarial examples, resulting in a high overall F1 score \cite{powers2011evaluation} of 96.39\%. Compared with other recent detection methods, our method can detect more adversarial examples with fewer false positives. In the defense-aware scenario, the experiments show that our detection method can effectively raise the bar for adversaries, the attack success rate is remarkably downgraded to 67.37\%. As demonstrated in \cite{carlini2017adversarial}, many recent detection methods are defeated by defense-aware attacks. In contrast, the proposed method is proven to be a promising mitigation to defense-aware attacks.

In summary, our main contributions are the following.
\begin{itemize}
\item
We model the perturbation of the DNN adversarial examples as image noise and introduce classic image processing techniques to reduce its effect. This allow us to effectively detect adversarial examples without prior knowledge of attack techniques.
\vspace{1pt}
\item
We employ the entropy to automatically adjust the detection strategy for a specific sample. This makes the proposed method capable of detecting different kinds of adversarial examples without requiring tuning its parameters, and can be directly integrated into unmodified target models.
\vspace{1pt}
\item
Using some state-of-the-art DNN models, we demonstrate that the proposed method can effectively detect the adversarial examples generated by different attack techniques with a better performance than other recent detection method, especially in the defense-aware attack scenario.\footnote{The source code of our detection method, along with the experiment data, is all available at \url{https://github.com/OwenSec/DeepDetector.}}
\end{itemize}
\section{Background}\label{sec:Background}
In this section, we provide some preliminaries on attack scenarios and the attack techniques used in our experiments. 

\subsection{Attack Scenarios}
In practice, the adversaries may possess different levels of knowledge of target models. Accordingly, they may adopt different methodologies when launching attacks. Similarly as done in \cite{carlini2017adversarial,xu2017feature,Biggio2017Evasion}, we  evaluate our detection method in two different attack scenarios, i.e., \textit{defense-unaware} and \textit{defense-aware}.

\textbf{Defense-unaware Scenario.} As discussed in \cite{Liu2017Trojaning}, more and more well-trained DNN models can be obtained from online markets (e.g., Caffe Model Zoo). Some users may directly employ such public models in their tasks \cite{esteva2017dermatologist,singla2016food}. For example, Esteva et al. \cite{esteva2017dermatologist} utilized a pre-trained Inception v3 model and retrained it on some clinical images to classify skin cancers. Both the original model and the retraining samples are publicly available. In other words, the adversary can also directly get the target model or build a same one and thoroughly analyze it. However, as illustrated in Fig. \ref{Fig2}, the proposed technique can be easily deployed on an unchanged trained model. The key component of the proposed technique is a filter \textit{T}, which can be directly integrated with an original model \textit{C} without requiring any modification to it. In fact, the user can stealthily introduce our detection technique in a public model. In this scenario, the adversary can completely understand \textit{C} but has no idea about the existence of \textit{T} (i.e., \textit{defense-unaware}).

\textbf{Defense-aware Scenario.} In some extreme cases, the adversaries can by all means get the whole target model, including the integrated detection mechanism. Consequently, they can fully analyze both \textit{C} and \textit{T} to generate desired adversarial examples and launch more sophisticated attacks.

\subsection{Crafting Adversarial Example}
Szegedy et al.~\cite{szegedy2013intriguing} first made the intriguing discovery that various machine learning models, including DNNs~\cite{le2013building}, are vulnerable to adversarial examples. In general, for a given sample \textit{x} and a trained model \textit{C}, the attacker aims to craft an adversarial example \textit{x$^*$} = \textit{x} + $\Delta$\textit{x} by adding a perturbation $\Delta$\textit{x} to \textit{x}, such that \textit{C}(\textit{x*}) $\neq$ \textit{C}(\textit{x}).

In most of the cases, the attacker wants the target model to misclassify the resultant image, while a human observer can still correctly classify it without noticing the existence of the introduced perturbation. In practice, the adversarial examples can be generated straightforwardly~\cite{goodfellow2014explaining} or with an optimization procedure~\cite{carlini2016towards,moosavi2016deepfool,szegedy2013intriguing}. In this paper, we choose the following three up-to-date attack techniques to perform detection experiments. They can produce imperceptible perturbations.

\textbf{Fast Gradient Sign Method}. Goodfellow et al.~\cite{goodfellow2014explaining} proposed a straightforward strategy named fast gradient sign method (FGSM) to craft adversarial examples against GoogLeNet~\cite{szegedy2015going}. The method is easy to implement and can compute adversarial perturbations very efficiently. Let \textit{c} be the true class of \textit{x} and \textit{J} (\textit{C}, \textit{x}, \textit{c}) be the cost function used to train the DNN \textit{C}. The perturbation is computed as the sign of the model's cost function gradient, i.e.
\begin{equation}
\Delta\textit{x} = \varepsilon~sign(\bigtriangledown_\textit{x}~\textit{J}~ (\textit{C},~\textit{x},~\textit{c}))
\end{equation}
where \textit{$\varepsilon$} (ranging from 0.0 to 1.0) is set to be small enough to make $\Delta$\textit{x} undetectable. Choosing a small \textit{$\varepsilon$} can produce a well-disguised adversarial example. The change to the original image is difficult to be spotted by a human. On the contrary, a large \textit{$\varepsilon$} is likely to introduce noticeable perturbations but can get more adversarial examples when the original images are simple (e.g., handwritten digits).

\begin{figure}[tb]
\centering
\includegraphics{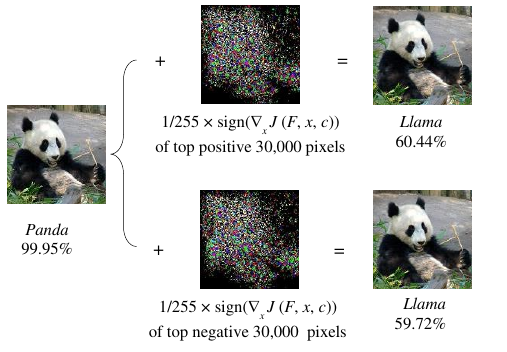}
\caption{Only manipulating the pixels with top gradients can still result in effectual adversarial examples.}
\label{Fig4}
\end{figure}

In the classic FGSM algorithm, all input pixels are applied either a positive or negative change in the same degree according to the direction (sign) of corresponding cost gradients. However, as illustrated in Fig. \ref{Fig4}, we found that only manipulating the 30,000 (19.92\%) input pixels with the highest positive or negative gradient magnitude can also generate an effectual adversarial example using the same \textit{$\varepsilon$}. The result implies that we can't assume the perturbation follows some kind of distribution.

\textbf{DeepFool}.  Moosavi-Dezfooli et al.~\cite{moosavi2016deepfool} devised the DeepFool algorithm to find very small perturbations that are sufficient to change the classification result. The original image \textit{x}$_0$ is manipulated iteratively. At each iteration of the algorithm, the perturbation vector for \textit{x}$_\textit{i}$ that reaches the decision boundary is computed, and the current estimate is updated. The algorithm stops until the predicted class of \textit{x}$_\textit{i}$ changes. DeepFool is implemented as an optimization procedure which can yield a good approximation of the minimal perturbation. Moosavi-Dezfooli et al. performed some attack experiments against several DNN image classifiers, such as GoogLeNet~\cite{szegedy2015going} and CaffeNet~\cite{jia2014caffe}, and so on. Their experiments demonstrated that DeepFool can lead to a smaller perturbation than FGSM, which however is still effective to trick the target models.

\textbf{CW Attacks}. Carlini and Wagner~\cite{carlini2016towards} also employed an optimization algorithm to seek as small as possible perturbations. Three powerful attacks (CW attacks for short) are designed for the \textit{L}$_0$, \textit{L}$_2$, and \textit{L}$_\infty$ distance metrics. Using some public datasets, such as MNIST~\cite{URL:TheMNISTDatabaseOfHandwrittenDigits} and ImageNet~\cite{deng2009imagenet}, they trained some deep network models to evaluate their attack methods. As demonstrated in~\cite{carlini2016towards}, CW attacks can find closer adversarial examples than the other attack techniques and never fail to find an adversarial example. For example, CW \textit{L}$_0$ and \textit{L}$_2$ attacks can find adversarial examples with at least 2 times lower distortion than FGSM. Besides, Carlini and Wagner also illustrated their attacks can effectively break the defensive distillation~\cite{papernot2016distillation}.
\section{Methodology}\label{sec:ApproachOverview}

\subsection{Overview}
The basic idea behind our method is to regard the perturbation as a kind of noise and introduce image processing techniques to reduce its adversarial effect as far as possible.

As described in Section 2, an adversarial example is crafted by superimposing some perturbations on the original image. In this sense, the perturbation introduced in the adversarial example is an additive noise item $\eta$(\textit{m}, \textit{n}), and the adversarial example can be considered as a degraded image \textit{g}(\textit{m}, \textit{n}) of the original image \textit{f}(\textit{m}, \textit{n}) as follows,
\begin{equation}
\textit{g}(\textit{m}, \textit{n}) = \textit{f}(\textit{m}, \textit{n}) + \eta(\textit{m}, \textit{n})
\end{equation}
where \textit{m} and \textit{n} are spatial coordinates, and \textit{f}, \textit{g} and $\eta$ are the functions mapping a pixel of coordinate (\textit{m}, \textit{n}) to its intensity. 

For example, the perturbation of an FGSM adversarial example can be viewed as an additive noise whose amplitude is $\varepsilon$. Ideally, if we can reconstruct the original image \textit{f}(\textit{m}, \textit{n}) from an adversarial example \textit{g}(\textit{m}, \textit{n}), adversarial examples can be detected immediately. However, achieving this is very difficult, if not impossible, due to lack of the necessary knowledge about the noise term $\eta$(\textit{m}, \textit{n}). Instead, we seek to reconstruct the original image in the sense of classification. Namely, we want to convert \textit{g}(\textit{m}, \textit{n}) to a new image \textit{f}'(\textit{m}, \textit{n}) such that its predicted class \textit{C}(\textit{f}'(\textit{m}, \textit{n})) can be the same as \textit{C}(\textit{f}(\textit{m}, \textit{n})).

\begin{figure}
\centering
\includegraphics{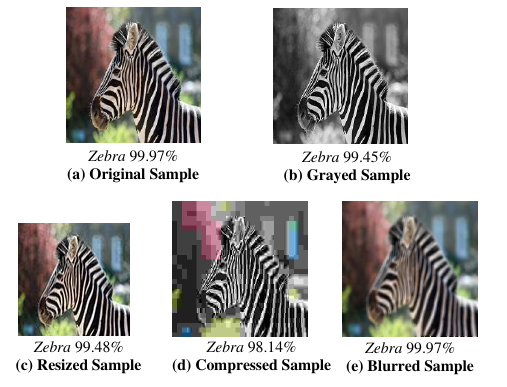}
\caption{GoogLeNet can correctly classify the processed images.}
\label{Fig5}
\end{figure}

Naturally, we hope that the model can correctly identify a benign sample after the conversion. If so, the adversarial example can be detected by checking whether the classification of a sample is changed. If the classification is changed, the sample will be identified as a potential adversarial example. Otherwise, it is considered benign. Fortunately, the state-of-the-art DNN image classifiers are robust to some extent and can tolerate a certain degree of distortion \cite{goodfellow2009measuring,lecun1989backpropagation}, although they are weak when facing adversarial examples. As Fig. \ref{Fig5} illustrates, the original image is classified as  \textit{Zebra} by GoogLeNet with 99.97\% confidence. After being either \textit{grayed}, \textit{resized}, \textit{compressed} or \textit{blurred}, it will still be correctly classified with high confidence.

As mentioned in Section 1, the noise reduction techniques are leveraged to reduce the effect of the perturbation. Based on the above discussions, we believe that although some details of interest in the sample may be removed too, the classifiers can output the correct classification for a denoised image. Note that, though the perturbation is treated as a kind of noise in this study, we cannot imply that the perturbation follows certain distributions. As presented in \cite{moosavi2016deepfool}, the perturbation may differ largely from one attack to another. This makes identifying them a nontrivial work. Moreover, in real-life scenarios, we are unlikely to know what technique the adversary uses and what features the perturbation may possess. In fact, the introduced perturbation can be random and there is not a predictable distribution about them, as illustrated in Fig. \ref{Fig4}. For this reason, rather than the prior knowledge-based techniques (e.g. \textit{Lee filtering}~\cite{lee1980digital}), we adopt two straightforward techniques that require no prior knowledge, namely \textit{scalar} \textit{quantization} and \textit{smoothing} \textit{spatial} \textit{filter}, to detect adversarial examples.

For scalar quantization, the size of intervals is a key parameter. In principle, using large intervals can more effectively reduce the effect of the perturbation but introduce more distortions at the same time, and the ``business'' of an image is damaged more heavily. This may result in a misclassification for a quantized benign sample, and produce a false positive. On the contrary, smaller intervals may bring more false negatives due to inadequate noise reduction.

We utilize the image \textit{entropy} to determine the parameter. The concept of image entropy describes how much randomness (or uncertainty) an image possesses, in other words, how much information is provided by the image \cite{tsai2008information}. The entropy value of an image is a quantitative measure of the information transmitted by the image. Commonly, the higher the entropy of an image is, the richer its semantics (``business'') often is. Consequently, for an image with higher entropy, more information is required for the classifier to identify its class. Based on the intuition, to avoid excessively eliminating the information of a sample, a small interval size will be applied to the high-entropy samples when quantizing them. Accordingly, the low-entropy samples will be assigned with a large interval size.

Smoothing a sample will blur its details and often decrease its information. However, for a very simple image (with a low entropy), e.g., a handwritten digit, the smoothing may excessively eliminate its details, which are important to the classification task. Namely, the low-entropy image can't tolerate the blurring well from the perspective of classification. To this end, we use the entropy to decide whether the sample needs to be smoothed.

\subsection{Evaluation Targets and Metrics}
\textbf{Target Models and Datasets.} According to the three attack techniques presented in Section 2, we do our best to collect five available target models to explore the effectiveness of the proposed method. As shown in Table \ref{models}, the models cover all three attack techniques and refer to two datasets, MNIST \cite{URL:TheMNISTDatabaseOfHandwrittenDigits} and ImageNet \cite{deng2009imagenet}. MNIST is a dataset of simple gray handwritten digits, while ImageNet is a large-scale dataset of labeled high-resolution images. We select the whole MINIST test dataset (10,000 digits) and seven classes of images (3,730 samples) of ImageNet for evaluation.

To prevent over-fitting, we determine hyper-parameters in a \textit{cross validation} way. As listed in Table \ref{datasets}, these samples are divided into three parts, \textit{training set}, \textit{validation set} and \textit{test set}. The training set is used to determine various detection parameters (e.g., the interval size, the entropy threshold, the choice of spatial smoothing filter, etc.), and the validation set is leveraged to verify whether the selected parameter setting is indeed effective. Besides, we only employ FGSM attack method in the parameter training phase, and the $\varepsilon$ value is fixed to 0.2 for MNIST and 1/255 for ImageNet. After the parameters are determined, we will evaluate the performance of our method against all the three attack techniques on the test set.

\begin{table}[t]
\renewcommand{\arraystretch}{1.2}
\begin{center}
\caption{Target models.}
\label{models}
\begin{tabular}{|c|c|c|}
\hline
\textbf{Model}&\textbf{Dataset}&\textbf{Attack}\\
\hline
M1 \cite{papernot2016cleverhans}&MNIST&FGSM\\
\hline
M2 \cite{URL:RobustEvasionAttacksAgainstNeuralNetworkToFindAdversarialExamples}&MNIST&CW\\
\hline
CaffeNet \cite{URL:bvlc}&ImageNet&DeepFool\\
\hline
GoogLeNet \cite{URL:bvlc}&ImageNet&FGSM, DeepFool\\
\hline
Inception v3 {\cite{URL:RobustEvasionAttacksAgainstNeuralNetworkToFindAdversarialExamples}}&ImageNet&CW\\
\hline
\end{tabular}
\end{center}
\end{table}

\begin{table}[t]
\renewcommand{\arraystretch}{1.2}
\begin{center}
\caption{Target datasets.}
\label{datasets}
\begin{tabular}{|c|c|c|c|}
\hline
\textbf{Dataset}&\textbf{Training}&\textbf{Validation}&\textbf{Test}\\
\hline
\multirow{2}{*}{\makecell[cc]{MNIST}}&\multirow{2}{*}{No. 0$\sim$4499}&\multirow{2}{*}{No. 4500$\sim$5499}&\multirow{2}{*}{No. 5500$\sim$9999}\\
&&&\\
\hline
\multirow{3}{*}{ImageNet}&Goldfish (648)&&Zebra (503)\\
&Pineapple (520)& Jellyfish (618)&Panda (501)\\
&Clock (455)&&Cab (485)\\
\hline
\end{tabular}
\end{center}
\end{table}

\textbf{Evaluation Metrics.} To explore some hyper-parameters of our method and evaluate the effectiveness, we adopt the \textit{recall} rate, the \textit{precision} rate and the \textit{F1} score \cite{powers2011evaluation} to quantify the detection performance, which are defined as follows,
\begin{equation}
\textit{Recall} = \frac{\textit{TP}}{\textit{TP} + \textit{FN}} \qquad
\end{equation}
\begin{equation}
\textit{Precision} = \frac{\textit{TP}}{\textit{TP} + \textit{FP}} \qquad
\end{equation}
\begin{equation}
\textit{F1} = 2*\frac{\textit{Recall}*\textit{Precision}}{\textit{Recall} + \textit{Precision}}
\qquad
\end{equation}
where \textit{TP} is the number of correctly detected adversarial examples (true positives), \textit{FN} the number of adversarial examples that survive from our detection (false negatives), and \textit{FP} the number of benign images that are detected as adversarial examples (false positives). The higher F1 score indicates the better overall detection performance. Ideally, if the method can detect all the adversarial examples without introducing any false positives, the F1 score will be 1.0. 

Besides, to evaluate the capability of recovering the original classification of adversarial examples via our detection filter, we introduce a new metric \textit{RTP} (short for \textit{Recovered True Positives}). RTP indicates the number of such samples, which are successfully detected as adversarial and its filtered version can be correctly classified as its original class.

\subsection{Computing Entropy}
Our preliminary experiments indicate that the input image itself is also a parameter of determining effectual detection scheme. In other words, the scheme that works well on detecting low-resolution adversarial images may fail detecting high-resolution ones. In this work, we leverage the discrete entropy to distinguish images with different resolutions. Discrete entropy \cite{gray2011entropy} is a powerful metric and has been proven useful in image processing \cite{wang2005brightness,yoo2012maximum,min2013novel}. Without loss of generality, for an M$\times$N grayscale image with 256 pixel levels (0$\sim$255), its entropy can be computed as equation (\ref{entropy}). The frequency of pixel level \textit{i} is denoted as $\textit{f}_{\textit{i}}$.
\begin{equation}
\textit{p}_{\textit{i}}=\textit{f}_{\textit{i}}~/~(M\times{N})~~~~~~~\textit{i} = 0,1,...,255
\end{equation}
\begin{equation}
H= -\sum_{\textit{i=0}}^{255}\textit{p}_{\textit{i}}log_2(\textit{p}_{\textit{i}})
\label{entropy}
\end{equation}
For a RGB color image, its entropy is the average of the entropies of its three color planes.

\begin{table}[t]
\renewcommand{\arraystretch}{1.2}
\caption{Uniform vs. non-uniform quantization.}
\label{table:quantization}
\centering
\begin{tabular}{|c|c|c|c|c|} 
\hline 
\textbf{Quantization} & \textbf{Time (s)} & \textbf{Recall} & \textbf{Precision} & \textbf{F1}\\
\hline 
Uniform&0.004&94.00\%&100.00\%&96.91\%\\
\hline
Non-uniform&137.7&94.00\%&51.65\%&66.67\%\\
\hline 
\end{tabular}
\end{table}

\begin{table*}[t]
\renewcommand{\arraystretch}{1.2}
\begin{center}
\caption{Performance of detecting FGSM adversarial examples with different scalar quantization schemes.}
\label{trainingResults}
\begin{tabular}{|c|c|c|c|c|c|c|c|c|c|c|}
\hline
\multirow{2}{*}{\textbf{Dataset}}&\multirow{2}{*}{\textbf{Metric}}&\multicolumn{9}{c|}{\textbf{Number of Intervals}}\\
\cline{3-11}
&&2&3&4&5&6&7&8&9&10\\
\hline
\multirow{3}{*}{MNIST}&Recall&\textbf{93.86\%}&86.48\%&93.04\%&13.26\%&7.25\%&24.95\%&42.03\%&50.62\%&52.35\%\\
\cline{2-11}
&Precision&\textbf{99.14\%}&99.66\%&99.90\%&99.51\%&99.11\%&99.74\%&99.77\%&100.00\%&100.00\%\\
\cline{2-11}
&F1 Score&\textbf{96.43\%}&92.60\%&96.35\%&23.40\%&13.51\%&39.92\%&59.15\%&67.22\%&68.72\%\\
\hline
\multirow{3}{*}{ImageNet}&Recall&94.35\%&93.07\%&91.26\%&86.59\%&\textbf{85.15\%}&80.71\%&76.56\%&68.50\%&66.84\%\\
\cline{2-11}
&Precision&57.51\%&72.14\%&82.05\%&88.32\%&\textbf{91.28\%}&93.54\%&94.60\%&94.79\%&95.27\%\\
\cline{2-11}
&F1 Score&71.46\%&81.28\%&86.41\%&87.44\%&\textbf{88.11\%}&86.65\%&84.63\%&79.53\%&78.57\%\\
\hline
\end{tabular}
\end{center}
\end{table*}

\subsection{Scalar Quantization}
\textit{Quantization} is the process of representing a large (possibly infinite) set of values with a smaller (finite) one, e.g., mapping the real numbers to the integers. In image processing,  quantization is often employed as a lossy compression technique by mapping a range of pixel intensities to a single representing one. In other words, reducing the number of colors of an image to cut its file size.

\textit{Scalar quantization} is the most practical and straightforward approach to quantize an image. In scalar quantization, all inputs within a specified interval are mapped to a common value (called codeword), and the inputs in a different interval will be mapped to a different codeword. There are two types of scalar quantization, \textit{uniform quantization} and \textit{non-uniform quantization} \cite{gonzalez2012digital}.  In uniform quantization, the input will be separated into the same size intervals, while in non-uniform quantization, the intervals are usually of different sizes chosen with an optimization algorithm to minimize the distortion~\cite{srivastava2012non,dziugaite2016study}. A small-scale empirical study is performed on 100 MNIST digits with FGSM attack to decide which type of quantization can more effectively downgrade the effect of the perturbation. We adopt the algorithm in \cite{srivastava2012non} to perform non-uniform quantization. The number of intervals is set to two for both uniform and non-uniform quantization. As Table \ref{table:quantization} shows, uniform quantization can achieve much higher F1 score in detection and spend much less time. Therefore, in this study, we employ the uniform quantization technique to handle the sample.

\begin{table}[t]
\renewcommand{\arraystretch}{1.2}
\caption{The denoising strategies for different entropies.}
\label{table1}
\centering
\begin{tabular}{|c|c|c|} 
\hline 
\textbf{Entropy} & \textbf{Quantization Intervals} & \textbf{Smoothing?}\\
\hline 
$<$4&2&NO\\
\hline 
4 $\sim$5& 4 & NO \\
\hline 
$>$5& 6 & YES\\
\hline 
\end{tabular}
\end{table}

\begin{table}[t]
\renewcommand{\arraystretch}{1.2}
\caption{Detection performance of the adaptive  quantization strategy.}
\label{val}
\begin{center}
\begin{tabular}{|c|c|c|c|c|c|c|} 
\hline 
\textbf{Dataset}&\textbf{TP}&\textbf{FN}&\textbf{FP}&\textbf{Recall}&\textbf{Precision}&\textbf{F1 Score}\\
\hline 
Training&3482&370&146&90.39\%&95.98\%&93.10\%\\
\hline
Validation&939&81&48&92.06\%&95.14\%&93.57\%\\
\hline
\end{tabular}
\end{center}
\end{table}

\begin{figure}[t]
\centering
\includegraphics[width=9cm,height=2.6cm]{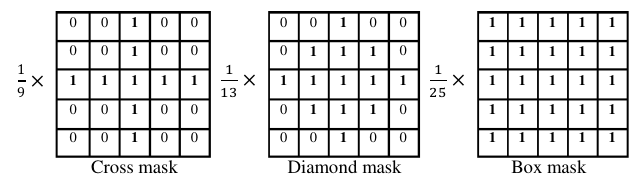}
\caption{Different filter masks with size $5\times5$.}
\label{filters}
\end{figure}

\begin{table*}[t]
\renewcommand{\arraystretch}{1.2}
\begin{center}
\caption{Performance of detecting high-entropy FGSM adversarial examples with different spatial smoothing filters.}
\label{trainfilters}
\begin{tabular}{|c|c|c|c|c|c|c|c|c|c|c|c|c|}
\hline
\multirow{2}{*}{\textbf{Filter Mask}}&\multicolumn{4}{c|}{\textbf{Cross Mask}}&\multicolumn{4}{c|}{\textbf{Diamond Mask}}&\multicolumn{4}{c|}{\textbf{Box Mask}}\\
\cline{2-13}
&$3\times3$&$\bm{5\times5}$&$\bm{7\times7}$&$9\times9$&$3\times3$&$\bm{5\times5}$&$\bm{7\times7}$&$9\times9$&$3\times3$&$\bm{5\times5}$&$7\times7$&$9\times9$\\
\hline
Recall&56.43\%&\textbf{84.05\%}&\textbf{90.03\%}&91.03\%&56.34\%&\textbf{85.54\%}&\textbf{90.93\%}&92.22\%&74.18\%&\textbf{91.53\%}&92.12\%&93.22\%\\
\hline
Precision&90.85\%&\textbf{91.83\%}&\textbf{86.00\%}&80.23\%&90.85\%&\textbf{88.27\%}&\textbf{81.57\%}&75.39\%&88.15\%&\textbf{82.41\%}&74.28\%&66.08\%\\
\hline
F1 Score&69.62\%&\textbf{87.77\%}&\textbf{87.97\%}&85.29\%&69.62\%&\textbf{86.89\%}&\textbf{86.00\%}&82.96\%&80.56\%&\textbf{86.73\%}&82.24\%&77.34\%\\
\hline
\end{tabular}
\end{center}
\end{table*}

\begin{table}[t]
\renewcommand{\arraystretch}{1.2}
\begin{center}
\caption{Performance of the five superior spatial smoothing filters with the validation set.}
\label{validationForFilters}
\begin{tabular}{|c|c|c|c|c|p{1.2cm}<\centering|}
\hline
\multirow{2}{*}{\textbf{Filter Mask}}&\multicolumn{2}{c|}{\textbf{Cross Mask}}&\multicolumn{2}{c|}{\textbf{Diamond Mask}}&$\!\!$\textbf{Box Mask}\\
\cline{2-6}
&$5\times5$&$7\times7$&$5\times5$&$7\times7$&$5\times5$\\
\hline
Recall&80.33\%&84.78\%&82.20\%&87.12\%&85.95\%\\
\hline
Precision&90.74\%&83.60\%&86.67\%&78.98\%&80.48\%\\
\hline
F1~Score&85.22\%&84.19\%&84.38\%&82.85\%&83.13\%\\
\hline
\end{tabular}
\end{center}
\end{table}

To develop a uniform quantization, we also need to determine the number of intervals. We tried to use the samples of the training set to discover an appropriate interval number. Unfortunately, from the experiment results shown in Table \ref{trainingResults}, we can see that there is not a general interval setting for different samples. The best F1 score 96.43\% for detecting MNIST adversarial examples is achieved under two-interval setting. However, for ImageNet examples, the best score 88.11\% is got when using six-interval. To tackle this problem, we leverage the image entropy described in Section 3.3 to automatically apply an adaptive interval size for a given sample. We conduct a series of experiments on the training set to seek some entropy thresholds, which will be used to set the interval size. Finally, as shown in Table \ref{table1}, we find a practicable solution. The number of intervals is set to two, four and six for the samples with different entropies (smaller than 4.0, between 4.0 and 5.0, and larger than 5.0) respectively. As Table \ref{val} presents, the cross validation experiments also show that the adaptive scalar quantization strategy can achieve satisfying detection performance on both training and validation set.

However, the experiments also show that there is still room to improve the detection performance on high-entropy (larger than 5.0) samples. The F1 score is 85.80\% for high-entropy samples but 95.68\% for the others. In essence, scalar quantization is a kind of \textit{point operation}. Subsequently, we further reduce perturbation for high-entropy samples by introducing the \textit{neighborhood operation} technique. As Table \ref{table1} presents, we also use the entropy to determine whether the smoothing should be performed.

\subsection{Spatial Smoothing Filter}
The \textit{spatial smoothing filter} is one of the most classic techniques for noise reduction. The idea behind it is to modify the value of the pixels in an image based on a local neighborhood of the pixels. As a result, the sharp transitions in pixel intensities, often brought by noise, are reduced in the target image. In linear smoothing filtering, the filtered image \textit{f}'(\textit{m}, \textit{n}) is the \textit{convolution} of the original image \textit{f}(\textit{m}, \textit{n}) with a filter mask \textit{w}(\textit{m}, \textit{n}) as follows.
\begin{equation}
 \textit{f}~'(\textit{m}, \textit{n}) = (\textit{w} * \textit{f}~)(\textit{m}, \textit{n}) = \sum_{\textit{s=-a}}^{\textit{a}}\sum_{\textit{t=-b}}^{\textit{b}}\textit{w}~(\textit{s}, \textit{t})~\textit{f}~(\textit{m}-\textit{s}, \textit{n}-\textit{t})
\end{equation} 

The filter mask determines the smoothing effect. Fig. \ref{filters} presents three common 5$\times$5 averaging filter masks. With a mask, the intensity of a pixel is replaced with the standard average of the intensities of the pixels which are weighted by 1 in its neighborhood. After filtered, the target image is blurred and small details are removed from it. However, from the viewpoint of image classification, the objects of interest may be highlighted and easy to detect.

To determine a practicable filter, we conduct a number of experiments with different shape and size of masks on the 2,015 high-entropy training samples. From the results listed in Table \ref{trainfilters}, it can be concluded that there are five acceptable candidates, i.e., $5\times5$ cross, $7\times7$ cross, $5\times5$ diamond, $7\times7$ diamond, and $5\times5$ box masks. The cross validation experiments (see Table \ref{validationForFilters}) also indicate that these five masks are effective on our validation set. In the next subsection, we will present the complete detection filter based on scalar quantization and spatial smoothing.

\subsection{Detection Filter}

In practice, the smoothing may bring an excessive blurring to some samples and produce new false positives. To this end, we design a combination filter based on the two above techniques rather than simply concatenating them together.

As discussed in Section 2, the attacker often wants the perturbation introduced in the adversarial example as small as possible to make it imperceptible. In other words, the perturbation to the pixel intensity is often limited in a small range. If the intensity of a pixel is blurred too much by the smoothing filter, the smoothing might be unnecessary. Based on the above intuition, our detection filter for high-entropy samples is defined by the following equation,
\vspace{1pt}
\textit{f}'(\textit{m},\textit{n}) = 
\begin{equation}
\left\{
\begin{array}{lr}
\textit{f}_{SQ}(\textit{m},\textit{n}),~\textbf{if} \mid f_{SQ}(\textit{m}, \textit{n})-\textit{f}(\textit{m},\textit{n})\mid\\ ~~~~~~~~~~~~~~~~~~~~~~~~~~~~~~~\leq ~ \mid \textit{f}_{SQ-SF}(\textit{m},\textit{n})-\textit{f}(\textit{m},\textit{n})\mid\\
\textit{f}_{SQ-SF}(\textit{m}, \textit{n}),~\textbf{else}
\end{array}
\right.
\label{detectionFilter}
\end{equation}
where \textit{f}$_{SQ}$(\textit{m}, \textit{n}) is the quantized original image and \textit{f}$_{SQ-SF}$(\textit{m}, \textit{n}) is the smoothed quantized image. For a given pixel (\textit{a}, \textit{b}) of the input sample \textit{f}(\textit{m}, \textit{n}), the output pixel value \textit{f}'(\textit{a}, \textit{b}) will be replaced with its quantization \textit{f}$_{SQ}$(\textit{a}, \textit{b}) when the distance between the quantization \textit{f}$_{SQ}$(\textit{a}, \textit{b}) and the original pixel value \textit{f}(\textit{a}, \textit{b}) is smaller than the one between \textit{f}$_{SQ-SF}$(\textit{a}, \textit{b}) and \textit{f}(\textit{a}, \textit{b}); otherwise, it will be set to \textit{f}$_{SQ-SF}$(\textit{a}, \textit{b}).

The complete framework of our detection method is ready now. In the first place, the image entropy is leveraged to automatically adjust detection parameters for each sample. For low-entropy (smaller than 4.0) and mid-entropy (between 4.0 and 5.0) samples, we only apply two-interval and four-interval uniform quantization to them respectively; for high-entropy (larger than 5.0) samples, we first quantize them with six-interval quantization, and then smooth their quantized version as illustrated in Equation \ref{detectionFilter}. To determine the final smoothing filter from the five candidates, we conduct a series of experiments and find that $7\times7$ cross mask is the best one when integrating it with scalar quantization. As Table \ref{valPreformance} shows, the final detection filter can achieve satisfying detection performance on both training and validation set.

Note that the proposed method is transparent to the target model. In practice, the detection filter can be directly integrated with any off-the-shelf model as a sample preprocessor. The target model can be kept unchanged.

\begin{table}[t]
\renewcommand{\arraystretch}{1.2}
\caption{Performance of the proposed detection filter on detecting FGSM adversarial examples.}
\label{valPreformance}
\begin{center}
\begin{tabular}{|c|c|c|c|c|c|c|} 
\hline 
\textbf{Dataset}&\textbf{TP}&\textbf{FN}&\textbf{FP}&\textbf{Recall}&\textbf{Precision}&\textbf{F1}\\
\hline 
Training&3324&266&108&92.59\%&96.85\%&94.67\%\\
\hline
Validation&1028&61&35&94.40\%&96.71\%&95.54\%\\
\hline
\end{tabular}
\end{center}
\end{table}
\section{Evaluation}\label{sec:Evaluation}

In this section, we evaluate the proposed method on our test set, consisting of 4,500 MNIST digits and 1,489 ImageNet images of three classes (\textit{Zebra}, \textit{Panda} and \textit{Cab}). Additionally, all the three attack techniques, i.e., FGSM, DeepFool and CW, are employed to inspect the effectiveness of our method. It should be pointed out that all the test samples and the latter two attack techniques (DeepFool and CW) are not involved in the design phase of our method. As mentioned in Section 1, the proposed method will be evaluated in in both defense-unaware and defense-aware scenarios.

\begin{table*}[t]
\renewcommand{\arraystretch}{1.2}
\begin{center}
\caption{Detection results of the proposed method. \#F denotes the number of images that cannot be perturbed to adversarial ones.}
\label{testResults}
\begin{tabular}{|c|c|c|c|c|c|c|c|c|c|c|c|}
\hline
\textbf{No.}&\textbf{Attack/Model}&\textbf{Dataset}&\textbf{$\#$F}&\textbf{TP}&\textbf{FN}&\textbf{FP}&\textbf{RTP}&\textbf{RTP\%}&\textbf{Recall}&\textbf{Precision}&\textbf{F1}\\
\hline
1&FGSM ($\varepsilon$=0.1)/M1&MNIST&4026&410&40&24&398&97.07\%&91.11\%&94.47\%&92.76\%\\
\hline
2&FGSM ($\varepsilon$=0.2)/M1&MNIST&1910&2467&106&32&2430&98.50\%&95.88\%&98.72\%&97.28\%\\
\hline
3&FGSM ($\varepsilon$=0.3)/M1&MNIST&455&3856&172&32&3768&97.71\%&95.73\%&99.18\%&97.42\%\\
\hline
4&FGSM ($\varepsilon$=0.4)/M1&MNIST&132&4078&273&32&3820&93.67\%&93.73\%&99.22\%&96.40\%\\
\hline
5&FGSM ($\varepsilon$=1/255)/GoogLeNet&ImageNet&270&841&98&88&718&85.37\%&89.56\%&90.53\%&90.04\%\\
\hline
6&FGSM ($\varepsilon$=2/255)/GoogLeNet&ImageNet&119&860&230&89&647&75.23\%&78.90\%&90.62\%&84.36\%\\
\hline
7&DeepFool/GoogLeNet&ImageNet&405&725&42&53&690&95.17\%&94.52\%&93.19\%&93.85\%\\
\hline
8&DeepFool/CaffeNet&ImageNet&96&900&29&47&854&94.89\%&96.88\%&95.04\%&95.95\%\\
\hline
\textbf{9}&\textbf{CW \textit{L}$_2$ ($\kappa$=0.0)/M2}&\textbf{MNIST}&\textbf{0}&\textbf{984}&\textbf{11}&\textbf{9}&\textbf{919}&\textbf{93.39\%}&\textbf{98.89\%}&\textbf{99.09\%}&\textbf{98.99\%}\\
\hline
10&CW \textit{L}$_2$ ($\kappa$=0.5)/M2&MNIST&0&984&11&9&920&93.50\%&98.89\%&99.09\%&98.99\%\\
\hline
11&CW \textit{L}$_2$ ($\kappa$=1.0)/M2&MNIST&0&983&12&9&913&92.88\%&98.79\%&99.09\%&98.94\%\\
\hline
12&CW \textit{L}$_2$ ($\kappa$=2.0)/M2&MNIST&0&979&16&9&897&91.62\%&98.39\%&99.09\%&98.74\%\\
\hline
13&CW \textit{L}$_2$ ($\kappa$=4.0)/M2&MNIST&0&959&36&9&866&90.30\%&96.38\%&99.07\%&97.71\%\\
\hline
14&CW \textit{L}$_2$ ($\kappa$=0.0)/Inception v3&ImageNet&2&98&0&8&91&92.86\%&100\%&92.45\%&96.08\%\\
\hline
15&CW \textit{L}$_2$ ($\kappa$=0.5)/Inception v3&ImageNet&0&100&0&2&98&98.00\%&100\%&98.04\%&99.01.\%\\
\hline
16&CW \textit{L}$_2$ ($\kappa$=1.0)/Inception v3&ImageNet&0&100&0&2&98&98.00\%&100\%&98.04\%&99.01.\%\\
\hline
17&CW \textit{L}$_2$ ($\kappa$=2.0)/Inception v3&ImageNet&0&100&0&2&98&98.00\%&100\%&98.04\%&99.01.\%\\
\hline
18&CW \textit{L}$_2$ ($\kappa$=4.0)/Inception v3&ImageNet&0&100&0&2&98&98.00\%&100\%&98.04\%&99.01.\%\\
\hline
19&CW \textit{L}$_\infty$/M2&MNIST&0&991&7&1&991&100\%&99.30\%&99.90\%&99.60\%\\
\hline
20&CW \textit{L}$_\infty$/Inception v3&ImageNet&25&75&0&2&73&97.33\%&100.00\%&97.40\%&98.68\%\\
\hline
\multicolumn{3}{|c|}{Total/Average}&7440&20590&1083&461&19387&94.16\%&95.00\%&97.81\%&96.39\%\\
\hline
\end{tabular}
\end{center}
\end{table*}

\subsection{Defense-unaware Attack Detection}
In the defense-unaware scenario, the adversary is ignorant of the proposed method, and can only generate adversarial examples on the original model.

\subsubsection{Detecting FGSM Examples}
For FGSM attack, the $\varepsilon$ value is an important and adjustable parameter. In general, the larger the $\varepsilon$ is, the more adversarial examples FGSM can successfully crafted. However, an excessive $\varepsilon$ is likely to introduce noticeable perturbation and be easily spotted by a human. To evaluate the capability of our method for detecting adversarial examples with different magnitudes of perturbations, the adversarial examples are crafted with some acceptable $\varepsilon$ values. It is set between 0.1 to 0.4 for MNIST digits and 1/255 or 2/255 for ImageNet images. As listed in Table \ref{testResults} (rows 1$\sim$6), the proposed method can achieve an average F1 score of 95.37\% in detecting all FGSM examples. Surprisingly, the classifications of 94.16\% adversarial images can be correctly recovered by the detection filter. Note that, the detection performance is obtained without using any prior knowledge about adversarial examples.

\subsubsection{Detecting DeepFool Examples}
We use the algorithm provided in~\cite{URL:ASimpleAndAccurateMethodToFoolDeepNeuralNetworks} to generate DeepFool examples against GoogLeNet and CaffeNet\cite{URL:bvlc}. The detection performance is presented in Table \ref{testResults} (rows 7 and 8). Although DeepFool can produce a smaller perturbation than FGSM, the proposed method is still effective. Specifically, the proposed method can achieve high F1 scores of 93.85\% and 95.95\% with respect to GoogLeNet and CaffeNet, respectively. Besides, the classifications of 95.02\% DeepFool examples can be successfully recovered by our filter.

\subsubsection{Detecting CW Examples}
Among the three techniques, CW attack is the strongest one. It is an optimization-based algorithm which can seek out as small as possible perturbations. Carlini and Wagner~\cite{carlini2016towards} demonstrated that CW attack can find closer adversarial examples than the other attack techniques. For example, CW \textit{L}$_2$ attack can find adversarial examples with at least two times lower distortion than FGSM. Besides, they proved that CW attack can craft adversarial examples with very high success rate. The \textit{\#F} column of Table \ref{testResults} further demonstrates the statement.

In the experiments, we employ CW \textit{L$_2$} and \textit{L$_\infty$} attack to generate well-disguised adversarial examples to evaluate our method. However, CW attack is much more expensive than other attack techniques. Generating an ImageNet adversarial example with \textit{L}$_2$ and \textit{L}$_\infty$ can take about 5 and 40 minutes respectively. For this reason, we only picked the last 1,000 MNIST digits and 100 randomly selected ImageNet images from the test set as experiment dataset. 

CW \textit{L$_2$} attack can generate adversarial examples under various confidences. To comprehensively evaluate our method, when generating \textit{L$_2$} examples, we use different $\kappa$ values \cite{carlini2016towards} to control the confidence of generated examples. When $\kappa$ = 0.0, the adversarial examples will be with low-confidence. As $\kappa$ increases, the confidence will increase accordingly. In the experiments, we set $\kappa$ to 0.0, 0.5, 1.0, 2.0 and 4.0, and generate \textit{L$_2$} examples with confidence from about 50\% to over 90\%. As listed in Table \ref{testResults} (rows 9$\sim$20), our method illustrates excellent detection performance. It can achieve an average F1 score of 98.80\%. In addition, the classifications of 93.94\% CW examples can be recovered successfully. By the way, detecting a sample with the proposed method only takes less than one second. The introduced overhead is totally negligible compared with the time consumption of generating a CW example.

\subsubsection{The Proposed Method vs. Other Detections}

\begin{table}[t]
\renewcommand{\arraystretch}{1.2}
\caption{The proposed method vs. other detections.}
\label{vs}
\begin{center}
\begin{tabular}{|c|c|c|} 
\hline 
\multirow{2}{*}{\textbf{Method}}&\textbf{Defense-unaware}&\textbf{Defense-aware}\\
\cline{2-3} 
&\textbf{Detection Performance}&\textbf{Attack Success?}\\
\hline
MMD \cite{grosse2017statistical}&``fails to detect'' \cite{carlini2017adversarial}&-\\
\hline
Cascade&\multirow{2}{*}{92.00\% false positive}&\multirow{2}{*}{-}\\
 Classifier \cite{li2016adversarial}&&\\
\hline
Network&\multirow{2}{*}{75.00\% recall}&\multirow{2}{*}{98.00\%}\\
Uncertainty \cite{feinman2017detecting}&&\\
\hline
$3\times3$ Filter \cite{li2016adversarial}&80.00\% recall&YES\\
\hline
KDE \cite{feinman2017detecting}&``able to detect'' \cite{carlini2017adversarial}&YES\\
\hline 
PCA \cite{hendrycks2017early}&``does detect'' \cite{carlini2017adversarial}&YES\\
\hline
Dimensionality&\multirow{2}{*}{97.00\% recall}&\multirow{2}{*}{YES}\\
Reduction \cite{bhagoji2017dimensionality}&&\\
\hline
Adversarial&\multirow{2}{*}{98.00\% recall}&\multirow{2}{*}{100\%}\\
Training \cite{gong2017adversarial}&&\\
\hline
Adversarial&\multirow{2}{*}{98.50\% recall}&\multirow{2}{*}{100\%}\\
Retraining \cite{grosse2017statistical}&&\\
\hline
\textbf{The proposed}&\textbf{98.89\% recall}&\multirow{2}{*}{\textbf{67.37\%}}\\
\textbf{method}&\textbf{0.90\% false positive}&\\
\hline
\end{tabular}
\end{center}
\end{table}

Recently, Carlini and Wagner \cite{carlini2017adversarial} presented a very comprehensive survey on the effectiveness of detection methods. They collect ten recent detection methods they could find and re-implemented seven of them, which didn't release any source code. The CW \textit{L$_2$} attack algorithm was employed to craft adversarial examples to test the methods under different attack scenarios. Their study involved a great amount of work and provided a solid basis for estimate our detection method. In this study, we follow the methodology of Carlini and Wagner \cite{carlini2017adversarial} and leverage their result to compare our method with related ones in the defense-unaware and defense-aware scenarios.

One \cite{metzen2017detecting} of the ten methods is other type of detection and beyond the scope of this paper. All the rest nine ones are used in the comparison. As listed in the second column of Table \ref{vs}, five target detection methods \cite{feinman2017detecting,hendrycks2017early,bhagoji2017dimensionality, gong2017adversarial, grosse2017statistical} can effectively detect adversarial examples under defense-unaware scenario. Carlini and Wagner \cite{carlini2017adversarial} provided exact recall rates for three of them \cite{bhagoji2017dimensionality, gong2017adversarial, grosse2017statistical}. However, the other two target methods \cite{feinman2017detecting,li2016adversarial} are less effective and cannot achieve a high recall rate. The remaining two \cite{li2016adversarial, grosse2017statistical} have been proven to be ineffective when facing CW \textit{L$_2$} attack.

We adopt the same configurations as \cite{carlini2017adversarial}, i.e., using the default $\kappa$ value setting (0.0), to generate \textit{L$_2$} examples. The detection performance of our method is presented in Table \ref{testResults} (row 9). Our method can achieve a high recall rate of 98.89\% and a low  false positive rate of 0.90\%. Compared with the existing methods, we can conclude that our method outperforms them in the defense-unaware scenario.

\subsection{Defense-aware Attack detection}
In the defense-aware scenario, the adversary knows the technique details of the detection method. When launching an attack, he or she can leverage the knowledge to fool the classifier as well as the detector. As done in \cite{carlini2017adversarial}, we launch a defense-aware CW \textit{L$_2$} attack on 1,000 MNIST test samples.

Specifically, to evade our detection, the adversary needs to generate an adversarial example \textit{x}, which satisfies the constraint \textit{C}(\textit{x}) == \textit{C}(\textit{T}(\textit{x})). We modify the original CW \textit{L$_2$} algorithm to get such examples. The constraint is introduced in the iteration of seeking an effectual adversarial example. This ensures the output image of every optimization and its denoised version are classified as the same thing. If the algorithm can output an adversarial example successfully, it will evade our detection method. The experiments show that there are 32.63\% of the samples cannot be perturbed successfully to the target model equipped with our detection. In other words, the attack success rate is 67.37\%.

Carlini and Wagner \cite{carlini2017adversarial} also presented defense-aware attack experiments for the seven detection methods except the two that already fail in defense-unaware attacks. As done in Section 4.1.4, we also use their results to make a comparison with other methods. From the third column of Table \ref{vs}, we can see that three methods \cite{feinman2017detecting,grosse2017statistical,gong2017adversarial} become completely ineffective. An adversary can successfully generate adversarial examples for almost all of samples. And all the other four methods \cite{li2016adversarial, feinman2017detecting,  bhagoji2017dimensionality, gong2017adversarial} are also "broken" \cite{carlini2017adversarial}. By comparison, our method can remarkably downgrade the success rate from nearly 100\% to 67.37\%.

As discussed in \cite{carlini2016towards} and \cite{carlini2017adversarial}, it is still an open problem to construct defenses to defense-aware attacks. Based on the above experiments, we believe that our method effectively raises the bar for adversaries to launch successful attacks.

\subsection{Summary}
All in all, 43,346 samples are used to evaluate our method in the defense-unaware experiments, half of them are adversarial and half are benign. In detecting those adversarial examples generated by the three attack techniques, we achieve an overall recall of 95.00\% and an overall precision of 97.81\%, resulting in a high overall F1 score of 96.39\%. Note that our detection method inevitably leads to some false positives. For the 21,673 benign samples, our method causes 461 samples (2.13\%) to be misclassified, which is an acceptable performance degradation in security critical scenarios. In addition, the classifications of 94.16\% adversarial examples can be recovered by the proposed filter.

In practice, it is very difficult to effectively detect defense-aware attacks, as demonstrated in previous studies. All existing detection methods were broken by defense-aware CW attacks. As Section 4.2 shows, our method certainly raises the bar for adversaries, and is a promising mitigation to defense-aware attacks.

Furthermore, our method can be deployed directly to the target model without modifying or retraining it, and even without any prior knowledge of attack techniques. In fact, we only use FGSM algorithm for determining detection filters, but the selected filter can effectively defeat the other two much stronger attacks, i.e., DeepFool and CW.
\section{Discussion and Limitations}\label{sec:Discussion}

\begin{figure}
\centering
\includegraphics[height=3.5cm,width=8cm]{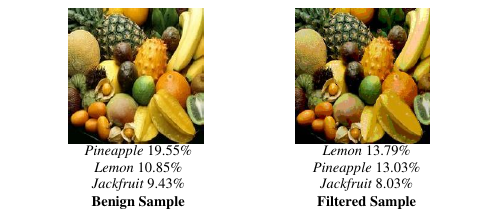}
\caption{A false positive caused by the ambiguous sample}
\label{Fig18}
\end{figure}

\textbf{False Positives and False Negatives.} In principle, the performance of our detection method is closely related to the classification capacity of target classifiers. Some false positives and false negatives are caused by the ambiguous images, which are essentially hard to classify for the target classifier. As shown in Fig. \ref{Fig18}, an image consisting of various fruits is labeled as \textit{Pineapple}, but GoogLeNet can tell that with only 19.55\% confidence. This is really not a strong prediction. The sample is also considered as a \textit{Lemon} with 10.85\% confidence and a \textit{Jackfruit} 9.43\%. After being denoised by our filter, the image is misclassified as \textit{Lemon} and results in a false positive. However, we think that neither the model nor the proposed detection filter is to blame for the false positive, but the ambiguity within the image is. The confusing images cannot only result in false positives, but also false negatives. Take an adversarial example generated with FGSM as example, the image shown in Fig. \ref{Fig19} is perturbed from \textit{Pineapple} to \textit{Sea Anemone} but only with 19.76\% confidence. GoogLeNet gives a weak prediction for it. And our detection method also fails to detect this adversarial example and produces a false negative.

There are quite a few ambiguous images like the two examples in our experiment set, which brings down the precision rate as well as the recall rate. For the ambiguous samples, inspecting more predict classes (e.g. top five predictions) might be necessary rather than just the top one with the highest confidence.

{\textbf{Perceptible Perturbations.} Some attack techniques, such as CW \textit{L}$_0$~\cite{carlini2016towards} and \textit{Jacobian-based saliency map approach} (JSMA) \cite{papernot2016limitations}, may introduce the large-amplitude perturbation. Such attack techniques limit the number of altered pixels, but not the amplitude of pixels. Consequently, as illustrated in Fig. \ref{Fig20}, the obtained adversarial example may present easy-to-notice distortions. It can be easily spotted by a human. However, it can still be exploited to launch an effective attack when the human interaction is not in consideration. Some other adversarial images with perceptible perturbations can be found in \cite{papernot2016limitations}. In principle, it is very difficult to properly reduce the effect of the heavy perturbation only with the filtering technique without compromising the semantics of the original image. To develop an effective technique to detect this kind of example is beyond the scope of this paper but will be our future research.

\textbf{Other Image Processing Techniques.} There are a number of other image processing techniques in addition to the ones adopted in our method. Some of them may can be leveraged to further improve our detection method, such as \textit{R\'enyi entropy}~\cite{chen2013short}, \textit{image segmentation}~\cite{gonzalez2012digital}, etc. For example, we can segment an adversarial example into some regions according to the connectivity among pixels to find such a region that possesses as much as possible information. From it, we have a good chance to restore the correct classification when the perturbation is isolated in other regions. In this way, the adversarial example shown in Fig. \ref{Fig20} may can be detected. In the future, we plan to investigate other image processing techniques to develop a more sophisticated detection method, especially for detecting the adversarial example with large-amplitude perturbations.

\begin{figure}
\centering
\includegraphics[height=3.5cm,width=8cm]{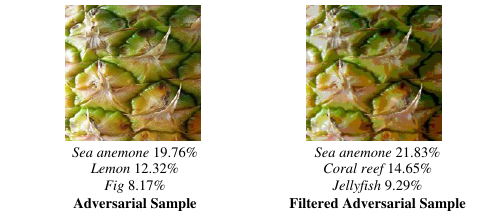}
\caption{A false negative caused by the ambiguous sample}
\label{Fig19}
\end{figure}

\begin{figure}
\centering
\includegraphics{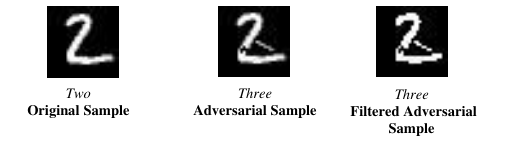}
\caption{A perceptible adversarial example generated with CW \textit{L}$_0$ attack.}
\label{Fig20}
\end{figure}
\section{Related Work}\label{sec:RelatedWork}

Many existing studies have paid much attention to the security of classifiers, and the arm race between adversaries and defenders will never end.

\textbf{Attacks on Traditional Classifiers}. Many studies have investigated the security of traditional machine learning methods~\cite{barreno2006can} and proposed some attack methods. Lowd and Meek conduct an attack that minimizes a cost function~\cite{lowd2005adversarial}. They further propose attacks against statistical spam filters that add the words indicative of non-spam emails to spam emails~\cite{lowd2005good}. The same strategy is employed in~\cite{nelson2008exploiting}. In ~\cite{maiorca2013looking}, a methodology, called \textit{reverse mimicry}, is designed to evade structural PDF malware detection systems. In ~\cite{laskov2014practical}, an online learning-based system, PDF$_\mathrm{RATE}$, was used as a case to investigate the effectiveness of evasion attacks. The study reconstructs a similar classifier through training one of the publicly available datasets by a few deduced features, and then evades PDF$_\mathrm{RATE}$ by insertion of dummy content into PDF files. In ~\cite{biggio2013evasion}, an algorithm is proposed for evasion of classifiers with differentiable discriminant functions. Liang et al. ~\cite{liang2016cracking} demonstrated that client-side classifiers are also vulnerable to evasion attacks.

Xu et al.~\cite{xu2016automatically} presented a general approach to find evasive variants by stochastically manipulate a malicious sample seed. The experiments showed that the effectual variants can be automatically generated to against two PDF malware classifiers, i.e.,  PDF$_\mathrm{RATE}$ and Hidost.

Fredrikson et al.~\cite{fredrikson2015model,wu2016methodology} developed a new form of model inversion attack which can infer sensitive features used in decision tree models and recover images from some facial recognition models by exploiting confidence values revealed by the target models. The proposed attack may cause serious privacy disclosure problems~\cite{fredrikson2015model}. More model inversion attacks can be found in~\cite{wu2016methodology}.

\textbf{Defenses for Traditional Classifiers}. Many countermeasures against evasion attacks have been proposed, such as using game theory~\cite{bruckner2011stackelberg,bruckner2012static} or probabilistic models~\cite{biggio2011design,rodrigues2009robustness} to predict attack strategy to construct more robust classifiers, employing multiple classifier systems (MCSs)~\cite{biggio2009multiple,biggio2010multiple,biggio2010multiple2} to increase the difficulty of evasion, and optimizing feature selection~\cite{globerson2006nightmare,kolcz2009feature} to make the features evenly distributed. 

Game-theoretical approaches~\cite{bruckner2011stackelberg,bruckner2012static} model the interactions between the adversary and the classifier as a game. The adversary's goal is to evade detection by minimally manipulating the attack instances, while the classifier is retrained to correctly classify them. 

MCSs~\cite{biggio2009multiple,biggio2010multiple,biggio2010multiple2}, as the name suggests, uses multiple classifiers rather than only one to improve classifier's robustness. The adversary who wants to effectively evade the classification has to fight with more than one classifier.

Kantchelian et al.~\cite{kantchelian2013approaches} present family-based ensembles of classifiers. In particular, they trained an ensemble of classifiers, one for each family of malware. By combining classifications, it will be determined whether an unknown binary is malware, and if it is, which family it belongs to. What's more, they also demonstrate the importance of human operators in adversarial environments.

In~\cite{globerson2006nightmare}, the method \textit{weight evenness} via feature selection optimization is proposed. By appropriate feature selection, the weight of every feature is evenly distributed, thus the adversary has to manipulate a larger number of features to evade detection. In~\cite{kolcz2009feature}, the features are reweighted inversely proportional to their corresponding importance, making it difficult for the adversary to exploit the features.

Unfortunately, these attack and defense techniques for traditional classifiers cannot be directly applied to DNNs. Along with the prevalence of DNNs, researchers have begun to pay close attention to the security of DNNs.

\textbf{Attacks on DNNs}. Recently, researchers have begun to attack DNN-based classifiers through crafting adversarial examples. There are various methods to generate adversarial examples against DNNs in various fields, not limited in computer vision~\cite{goodfellow2014explaining,moosavi2016deepfool,szegedy2013intriguing}, but also speech recognition~\cite{kereliuk2015deep}, text classification~\cite{liang2017deep} and malware detection~\cite{grosse2016adversarial}.

Kereliuk et al.~\cite{kereliuk2015deep} proposed a method to craft adversarial audio examples using the gradient information of the model's loss function. Through the application of minor perturbations to the input magnitude spectra, they can effectively craft an adversarial example. Text as discrete data is sensitive to perturbation. Liang et al.\cite{liang2017deep} proposed a method to craft adversarial text examples. Three perturbation strategies, namely \textit{insertion}, \textit{modification}, and
\textit{removal}, are designed to generate an adversarial example for a given text. By computing the cost gradients, what should be inserted, modified or removed, where to insert and how to modify are determined effectively. By elaborately dressing a text sample, the adversary can modify the classification to any other classes while still keeps the meaning unchanged. Grosse et al.~\cite{grosse2016adversarial} presented a method to craft adversarial examples on neural networks for malware classification, by adapting the method originally proposed in~\cite{papernot2016limitations}. 

In this paper, we focus on the detection of adversarial images. We believe that our method can be applied to detect adversarial examples for audio, which is also a kind of continuous data. But the proposed technique cannot be applied to discrete data, such as the adversarial text and malware. The new method need to be developed for effectively detecting them.

Note that there are two recent studies focus on crafting adversarial examples in the physical world. Kurakin et al.~\cite{kurakin2016adversarial} demonstrated that the adversarial images obtained from a cell-phone camera can still fool an ImageNet classifier. Sharif et al.~\cite{sharif2016accessorize} presented an attack method to fool facial biometric systems. They showed that with some well-crafted eyeglass frames, a subject can dodge recognition or impersonate others.

Besides, Shokri et al.~\cite{shokri2017Membership} developed a novel black-box membership inference attack against machine learning models, including DNN and non-DNN models. Given a data record, the attacker can determine whether it is in the target model's training dataset. For health-care datasets, such information leakage is unacceptable.

\textbf{Improve the Robustness of Deep Networks}. The adversarial training~\cite{goodfellow2014explaining,kereliuk2015deep,moosavi2016deepfool,papernot2016limitations,Kurakin2017Adversarial} is a straightforward defense technique to improve the robustness of target models. Retraining models by adding as many as possible adversarial examples can bring more challenges for attackers to find new adversarial examples. 

Wang et al.~\cite{wang2016learning} integrated a data transformation module right in front of a standard DNN to improve the model's resistance to adversarial examples. This data transformation module leverages non-parametric dimensionality reduction methods, and projects all the input samples into a new representation before passing the inputs to the target DNN in training and testing. Wang et al.~\cite{wang2016Adversary} also proposed another method, named \textit{random feature nullification}, for constructing adversary resistant DNNs. In particular, it randomly nullifies or masks features within input samples in both the training and testing phase. Such nullification makes a DNN model non-deterministic and then improves model's resistance to adversarial examples.

The proposed method is compatible with the above defense techniques. Defenders can still use our method in the enhanced model to get a better performance.

\textbf{Detection Techniques}. Some studies also focus on detecting adversarial examples directly.

Xu et al. in a very recent study~\cite{xu2017feature} proposed a method, called \textit{Feature Squeezing}, to detect adversarial examples in a similar way as ours. They explore two approaches to squeeze the features of an image: reducing the color bit depth of each pixel and smoothing it using a spatial filter. Their system identifies the adversarial examples by measuring the disagreement among the prediction vectors of the original and squeezed examples. Their experiments illustrated high performance was achieved when detecting FGSM adversarial examples in a MNIST model. However, a predefined threshold is required for determining how much disagreement indicates the current sample is adversarial. In their experiments, half of the examples are used to train the threshold that can produce the best detection accuracy on training examples. This means the defender must have a sufficient number of adversarial examples generated with potential attack techniques. As a result, the method works well only when the attack technique is known but less effective when facing unknown attacks. Moreover, in principle, for different datasets, models or attacks, the thresholds need to be retrained to achieve acceptable performance. By introducing the image entropy, we implement an adaptive detection method and can be directly applied to different models, datasets and attack techniques with the same setting and without requiring any prior knowledge of attacks.

Grosse et al.~\cite{grosse2017statistical} put forward a defense to detect adversarial examples using statistical tests. The method requires a sufficient large group of adversarial examples and benign examples to estimate their data distribution. However, the statistical test method cannot be directly applied to detect individual examples, making it less useful in practice. For this reason, Grosse et al. further propose a new method by adding an additional class (e.g., adversarial class) to the model's output and retraining the model to classify adversarial examples as the new class. Gong \textit{et al.} \cite{gong2017adversarial} proposed a similar adversarial retraining-based detection method. In the method, they trained a binary model to determine whether a given input is adversarial. Bhagoji et al. \cite{bhagoji2017dimensionality} leveraged the principle component analysis (PCA) technique to reduce the dimensionality of the images. Then instead of training on the original images, they trained a classifier on images which had been processed with dimensionality reduction. Hendrycks and Gimpel~\cite{hendrycks2017early} used a large number of adversarial examples to train a detector to identify unknown adversarial examples. A small "detector" subnetwork is trained on the binary classification task of distinguishing benign samples from adversarial perturbations. 

To a large extent, the performance of the above four detection techniques~\cite{grosse2017statistical,gong2017adversarial,hendrycks2017early,bhagoji2017dimensionality} also depends on how much effectual adversarial examples are available. 

Li \textit{et al.} \cite{li2016adversarial} built a cascade classifier and integrated the classifier with the original model. By checking each output of the original model's inner convolutional layers, they can determine whether a input sample is adversarial. Feinman et al.~\cite{feinman2017detecting} devised two novel features to detect adversarial examples based on the idea that adversarial examples deviate the true data manifold. They introduced density estimates to measure the distance between an unknown input sample and a set of benign samples. The method is computationally expensive and may be less effective in detecting adversarial examples which are very close to benign samples.

In summary, compared with the above defense techniques, our method possesses three advantages. First, our method is adaptive. The detection strategy is automatically adjusted for individual samples. The input image will be suitably filtered according to its entropy before being fed to the classifier. This improves the detection performance greatly. Second, the method is not attack-specific and thus holds great promise for detecting unknown attacks. As the experiments demonstrated, the detection parameters tuned with respect to a weaker attack (FGSM) is also applicable to stronger attacks (DeepFool and CW). Finally, our method is easier to deploy. The method can be directly integrated with a trained model, without retraining or modifying the model.
\section{Conclusion}\label{sec:Conclusion}

Many efforts have been paid to use various techniques to defend or detect the adversarial image examples in DNNs. However, the prior knowledge of attack techniques or the modification to the target model is often required. This paper presents a straightforward and effective adversarial image examples detection method. The adversarial perturbations are regarded as a kind of noise and the proposed method is implemented as a filter to reduce their effect. The image entropy is used to automatically adjust the detection strategy for specific samples. Our method provides two important features (1) without requiring the prior knowledge about attacks and (2) can be directly integrated into unmodified models.

We evaluate our method in both defense-unaware and defense-aware scenario. The experiments show that our method can achieve a high F1 score in detecting the adversarial examples generated by the different attack techniques and targeting different models in the defense-unaware scenario, and can effectively raise the bar for adversaries in the defense-aware scenario. Compared with existing detection methods, our method can perform better in both the two attack scenarios. Note that our method is also compatible with other defense techniques. A better performance can be achieved by combining them together.

Our research demonstrates that the adversarial images can be effectively analyzed with classic image processing techniques. In the future, we will investigate more image processing techniques to find more effective and practicable detection techniques, especially for defense-aware attacks.


%

\ifCLASSOPTIONcompsoc
  \section*{Acknowledgments}
\else
  \section*{Acknowledgment}
\fi

The authors would like to thank the anonymous reviewers for their insightful comments on the preliminary version of this paper. The work is supported by National Natural Science Foundation of China (NSFC) under grants 91418206, 61802413, 61672523, and 61472429, National Science and Technology Major Project of China under grant 2012ZX01039-004.

\ifCLASSOPTIONcaptionsoff
  \newpage
\fi



\bibliographystyle{IEEEtran}
\bibliography{DeepDetector}

\begin{thebibliography}{10}
\providecommand{\url}[1]{#1}
\csname url@samestyle\endcsname
\providecommand{\newblock}{\relax}
\providecommand{\bibinfo}[2]{#2}
\providecommand{\BIBentrySTDinterwordspacing}{\spaceskip=0pt\relax}
\providecommand{\BIBentryALTinterwordstretchfactor}{4}
\providecommand{\BIBentryALTinterwordspacing}{\spaceskip=\fontdimen2\font plus
\BIBentryALTinterwordstretchfactor\fontdimen3\font minus
  \fontdimen4\font\relax}
\providecommand{\BIBforeignlanguage}[2]{{%
\expandafter\ifx\csname l@#1\endcsname\relax
\typeout{** WARNING: IEEEtran.bst: No hyphenation pattern has been}%
\typeout{** loaded for the language `#1'. Using the pattern for}%
\typeout{** the default language instead.}%
\else
\language=\csname l@#1\endcsname
\fi
#2}}
\providecommand{\BIBdecl}{\relax}
\BIBdecl

\bibitem{hagan1996neural}
M.~T. Hagan, H.~B. Demuth, M.~H. Beale \emph{et~al.}, \emph{Neural network
  design}.\hskip 1em plus 0.5em minus 0.4em\relax Pws Pub. Boston, 1996,
  vol.~20.

\bibitem{liu2017survey}
W.~Liu, Z.~Wang, X.~Liu, N.~Zeng, Y.~Liu, and F.~E. Alsaadi, ``A survey of deep
  neural network architectures and their applications,'' \emph{Neurocomputing},
  vol. 234, pp. 11--26, 2017.

\bibitem{krizhevsky2012imagenet}
A.~Krizhevsky, I.~Sutskever, and G.~E. Hinton, ``Imagenet classification with
  deep convolutional neural networks,'' in \emph{Proceedings of Advances in
  Neural Information Processing Systems}, 2012, pp. 1097--1105.

\bibitem{lecun2010convolutional}
Y.~LeCun, K.~Kavukcuoglu, and C.~Farabet, ``Convolutional networks and
  applications in vision,'' in \emph{Proceedings of the 2010 IEEE International
  Symposium on Circuits and Systems (ISCAS)}.\hskip 1em plus 0.5em minus
  0.4em\relax IEEE, 2010, pp. 253--256.

\bibitem{dahl2012context}
G.~E. Dahl, D.~Yu, L.~Deng, and A.~Acero, ``Context-dependent pre-trained deep
  neural networks for large-vocabulary speech recognition,'' \emph{IEEE
  Transactions on Audio, Speech, and Language Processing}, vol.~20, no.~1, pp.
  30--42, 2012.

\bibitem{hinton2012deep}
G.~Hinton, L.~Deng, D.~Yu, G.~E. Dahl, A.-r. Mohamed, N.~Jaitly, A.~Senior,
  V.~Vanhoucke, P.~Nguyen, T.~N. Sainath \emph{et~al.}, ``Deep neural networks
  for acoustic modeling in speech recognition: The shared views of four
  research groups,'' \emph{IEEE Signal Processing Magazine}, vol.~29, no.~6,
  pp. 82--97, 2012.

\bibitem{collobert2008unified}
R.~Collobert and J.~Weston, ``A unified architecture for natural language
  processing: Deep neural networks with multitask learning,'' in
  \emph{Proceedings of the 25th international conference on Machine
  learning}.\hskip 1em plus 0.5em minus 0.4em\relax ACM, 2008, pp. 160--167.

\bibitem{zhang2015character}
X.~Zhang, J.~Zhao, and Y.~LeCun, ``Character-level convolutional networks for
  text classification,'' in \emph{Proceedings of Advances in Neural Information
  Processing Systems}, 2015, pp. 649--657.

\bibitem{szegedy2015going}
C.~Szegedy, W.~Liu, Y.~Jia, P.~Sermanet, S.~Reed, D.~Anguelov, D.~Erhan,
  V.~Vanhoucke, and A.~Rabinovich, ``Going deeper with convolutions,'' in
  \emph{Proceedings of the IEEE Conference on Computer Vision and Pattern
  Recognition}, 2015, pp. 1--9.

\bibitem{sun2014deep}
Y.~Sun, X.~Wang, and X.~Tang, ``Deep learning face representation from
  predicting 10,000 classes,'' in \emph{Proceedings of the IEEE Conference on
  Computer Vision and Pattern Recognition}, 2014, pp. 1891--1898.

\bibitem{sun2014deep2}
Y.~Sun, Y.~Chen, X.~Wang, and X.~Tang, ``Deep learning face representation by
  joint identification-verification,'' in \emph{Proceedings of Advances in
  Neural Information Processing Systems}, 2014, pp. 1988--1996.

\bibitem{goodfellow2014explaining}
I.~Goodfellow, J.~Shlens, and C.~Szegedy, ``Explaining and harnessing
  adversarial examples,'' in \emph{Proceedings of the 2015 International
  Conference on Learning Representations}, 2015.

\bibitem{moosavi2016deepfool}
S.-M. Moosavi-Dezfooli, A.~Fawzi, and P.~Frossard, ``Deepfool: a simple and
  accurate method to fool deep neural networks,'' in \emph{Proceedings of the
  IEEE Conference on Computer Vision and Pattern Recognition}, 2016, pp.
  2574--2582.

\bibitem{szegedy2013intriguing}
C.~Szegedy, W.~Zaremba, I.~Sutskever, J.~Bruna, D.~Erhan, I.~Goodfellow, and
  R.~Fergus, ``Intriguing properties of neural networks,'' in \emph{Proceedings
  of the 2014 International Conference on Learning Representations}, 2014.

\bibitem{papernot2016practical}
N.~Papernot, P.~McDaniel, I.~Goodfellow, S.~Jha, Z.~B. Celik, and A.~Swami,
  ``Practical black-box attacks against machine learning,'' in
  \emph{Proceedings of the 2017 ACM on Asia Conference on Computer and
  Communications Security}, 2017, pp. 506--519.

\bibitem{kereliuk2015deep}
C.~Kereliuk, B.~L. Sturm, and J.~Larsen, ``Deep learning and music
  adversaries,'' \emph{IEEE Transactions on Multimedia}, vol.~17, no.~11, pp.
  2059--2071, 2015.

\bibitem{papernot2016limitations}
N.~Papernot, P.~McDaniel, S.~Jha, M.~Fredrikson, Z.~B. Celik, and A.~Swami,
  ``The limitations of deep learning in adversarial settings,'' in
  \emph{Proceedings of the 2016 IEEE European Symposium on Security and Privacy
  (EuroS\&P)}.\hskip 1em plus 0.5em minus 0.4em\relax IEEE, 2016, pp. 372--387.

\bibitem{papernot2016distillation}
N.~Papernot, P.~McDaniel, X.~Wu, S.~Jha, and A.~Swami, ``Distillation as a
  defense to adversarial perturbations against deep neural networks,'' in
  \emph{Proceedings of the 2016 IEEE Symposium on Security and Privacy
  (S\&P)}.\hskip 1em plus 0.5em minus 0.4em\relax IEEE, 2016, pp. 582--597.

\bibitem{feinman2017detecting}
R.~Feinman, R.~R. Curtin, S.~Shintre, and A.~B. Gardner, ``Detecting
  adversarial samples from artifacts,'' \emph{arXiv preprint arXiv:1703.00410},
  2017.

\bibitem{metzen2017detecting}
J.~H. Metzen, T.~Genewein, V.~Fischer, and B.~Bischoff, ``On detecting
  adversarial perturbations,'' \emph{arXiv preprint arXiv:1702.04267}, 2017.

\bibitem{grosse2017statistical}
K.~Grosse, P.~Manoharan, N.~Papernot, M.~Backes, and P.~McDaniel, ``On the
  (statistical) detection of adversarial examples,'' \emph{arXiv preprint
  arXiv:1702.06280}, 2017.

\bibitem{xu2017feature}
W.~Xu, D.~Evans, and Y.~Qi, ``Feature squeezing: Detecting adversarial examples
  in deep neural networks,'' in \emph{Proceedings of Network and Distributed
  System Security Symposium}, 2018.

\bibitem{goodfellow2009measuring}
I.~Goodfellow, H.~Lee, Q.~V. Le, A.~Saxe, and A.~Y. Ng, ``Measuring invariances
  in deep networks,'' in \emph{Proceedings of Advances in Neural Information
  Processing Systems}, 2009, pp. 646--654.

\bibitem{URL:LeNet-5ConvolutionalNeuralNetworks}
``{LeNet-5, convolutional neural networks},''
  \url{http://yann.lecun.com/exdb/lenet/}.

\bibitem{lecun1989backpropagation}
Y.~LeCun, B.~Boser, J.~S. Denker, D.~Henderson, R.~E. Howard, W.~Hubbard, and
  L.~D. Jackel, ``Backpropagation applied to handwritten zip code
  recognition,'' \emph{Neural computation}, vol.~1, no.~4, pp. 541--551, 1989.

\bibitem{jia2014caffe}
Y.~Jia, E.~Shelhamer, J.~Donahue, S.~Karayev, J.~Long, R.~Girshick,
  S.~Guadarrama, and T.~Darrell, ``Caffe: Convolutional architecture for fast
  feature embedding,'' in \emph{Proceedings of the 22nd ACM international
  conference on Multimedia}.\hskip 1em plus 0.5em minus 0.4em\relax ACM, 2014,
  pp. 675--678.

\bibitem{URL:TheMNISTDatabaseOfHandwrittenDigits}
``{The MNIST database of handwritten digits},''
  \url{http://yann.lecun.com/exdb/mnist}.

\bibitem{deng2009imagenet}
D.~Jia, D.~Wei, S.~Richard, L.-J. Li, L.~Kai, and F.-F. Li, ``Imagenet: A
  large-scale hierarchical image database,'' in \emph{Proceedings of the IEEE
  Conference on Computer Vision and Pattern Recognition}.\hskip 1em plus 0.5em
  minus 0.4em\relax IEEE, 2009, pp. 248--255.

\bibitem{carlini2016towards}
N.~Carlini and D.~Wagner, ``Towards evaluating the robustness of neural
  networks,'' in \emph{Proceedings of the 2017 IEEE Symposium on Security and
  Privacy (S\&P)}.\hskip 1em plus 0.5em minus 0.4em\relax IEEE, 2017, pp.
  39--57.

\bibitem{carlini2017adversarial}
------, ``Adversarial examples are not easily detected: Bypassing ten detection
  methods,'' in \emph{Proceedings of the 10th ACM Workshop on Artificial
  Intelligence and Security}.\hskip 1em plus 0.5em minus 0.4em\relax ACM, 2017,
  pp. 3--14.

\bibitem{powers2011evaluation}
D.~M. Powers, ``Evaluation: from precision, recall and f-measure to roc,
  informedness, markedness and correlation,'' 2011.

\bibitem{le2013building}
Q.~V. Le, ``Building high-level features using large scale unsupervised
  learning,'' in \emph{Proceedings of the 2013 IEEE International Conference on
  Acoustics, Speech and Signal Processing (ICASSP)}.\hskip 1em plus 0.5em minus
  0.4em\relax IEEE, 2013, pp. 8595--8598.

\bibitem{lee1980digital}
J.-S. Lee, ``Digital image enhancement and noise filtering by use of local
  statistics,'' \emph{IEEE transactions on pattern analysis and machine
  intelligence}, no.~2, pp. 165--168, 1980.

\bibitem{tsai2008information}
D.-Y. Tsai, Y.~Lee, and E.~Matsuyama, ``Information entropy measure for
  evaluation of image quality,'' \emph{Journal of digital imaging}, vol.~21,
  no.~3, pp. 338--347, 2008.

\bibitem{papernot2016cleverhans}
N.~Papernot, I.~Goodfellow, R.~Sheatsley, R.~Feinman, and P.~McDaniel,
  ``cleverhans v1.0.0: an adversarial machine learning library,'' \emph{arXiv
  preprint arXiv:1610.00768}, 2016.

\bibitem{URL:RobustEvasionAttacksAgainstNeuralNetworkToFindAdversarialExamples}
``{Robust evasion attacks against neural network to find adversarial
  examples},'' \url{https://github.com/carlini/nn_robust_attacks}.

\bibitem{URL:bvlc}
``{BVLC models},'' \url{https://github.com/BVLC/caffe/tree/master/models}.

\bibitem{gray2011entropy}
R.~M. Gray, \emph{Entropy and information theory}.\hskip 1em plus 0.5em minus
  0.4em\relax Springer Science \& Business Media, 2011.

\bibitem{wang2005brightness}
C.~Wang and Z.~Ye, ``Brightness preserving histogram equalization with maximum
  entropy: a variational perspective,'' \emph{IEEE Transactions on Consumer
  Electronics}, vol.~51, no.~4, pp. 1326--1334, 2005.

\bibitem{yoo2012maximum}
J.-H. Yoo, S.-Y. Ohm, and M.-G. Chung, ``Maximum-entropy image enhancement
  using brightness mean and variance,'' \emph{Journal of Internet Computing and
  Services}, vol.~13, no.~3, pp. 61--73, 2012.

\bibitem{min2013novel}
B.~S. Min, D.~K. Lim, S.~J. Kim, and J.~H. Lee, ``A novel method of determining
  parameters of clahe based on image entropy,'' \emph{International Journal of
  Software Engineering and Its Applications}, vol.~7, no.~5, pp. 113--120,
  2013.

\bibitem{gonzalez2012digital}
R.~C. Gonzalez and R.~E. Woods, ``Digital image processing,'' 2012.

\bibitem{srivastava2012non}
M.~Srivastava and P.~K. Panigrahi, ``Non-uniform quantization of detail
  components in wavelet transformed image for lossy jpeg2000 compression,''
  \emph{arXiv preprint arXiv:1210.8165}, 2012.

\bibitem{dziugaite2016study}
G.~K. Dziugaite, Z.~Ghahramani, and D.~M. Roy, ``A study of the effect of jpg
  compression on adversarial images,'' \emph{arXiv preprint arXiv:1608.00853},
  2016.

\bibitem{URL:ASimpleAndAccurateMethodToFoolDeepNeuralNetworks}
``{A simple and accurate method to fool deep neural networks},''
  \url{https://github.com/lts4/deepfool}.

\bibitem{li2016adversarial}
X.~Li and F.~Li, ``Adversarial examples detection in deep networks with
  convolutional filter statistics,'' \emph{CoRR, abs/1612.07767}, vol.~7, 2016.

\bibitem{hendrycks2017early}
D.~Hendrycks and K.~Gimpel, ``Early methods for detecting adversarial images,''
  2017.

\bibitem{bhagoji2017dimensionality}
A.~N. Bhagoji, D.~Cullina, and P.~Mittal, ``Dimensionality reduction as a
  defense against evasion attacks on machine learning classifiers,''
  \emph{arXiv preprint arXiv:1704.02654}, 2017.

\bibitem{gong2017adversarial}
Z.~Gong, W.~Wang, and W.-S. Ku, ``Adversarial and clean data are not twins,''
  \emph{arXiv preprint arXiv:1704.04960}, 2017.

\bibitem{chen2013short}
B.~Chen and J.-j. Zhang, ``On short interval expansion of r{\'e}nyi entropy,''
  \emph{Journal of High Energy Physics}, vol.~11, p. 164, 2013.

\bibitem{barreno2006can}
M.~Barreno, B.~Nelson, R.~Sears, A.~D. Joseph, and J.~D. Tygar, ``Can machine
  learning be secure?'' in \emph{Proceedings of the 2006 ACM Symposium on
  Information, computer and communications security}.\hskip 1em plus 0.5em
  minus 0.4em\relax ACM, 2006, pp. 16--25.

\bibitem{lowd2005adversarial}
D.~Lowd and C.~Meek, ``Adversarial learning,'' in \emph{Proceedings of the
  eleventh ACM SIGKDD international conference on Knowledge discovery in data
  mining}.\hskip 1em plus 0.5em minus 0.4em\relax ACM, 2005, pp. 641--647.

\bibitem{lowd2005good}
------, ``Good word attacks on statistical spam filters.'' in \emph{Proceedings
  of the 2nd Conference on Email and Anti-Spam}, 2005.

\bibitem{nelson2008exploiting}
B.~Nelson, M.~Barreno, F.~J. Chi, A.~D. Joseph, B.~I. Rubinstein, U.~Saini,
  C.~A. Sutton, J.~D. Tygar, and K.~Xia, ``Exploiting machine learning to
  subvert your spam filter.'' \emph{LEET}, vol.~8, pp. 1--9, 2008.

\bibitem{maiorca2013looking}
D.~Maiorca, I.~Corona, and G.~Giacinto, ``Looking at the bag is not enough to
  find the bomb: an evasion of structural methods for malicious pdf files
  detection,'' in \emph{Proceedings of the 8th ACM SIGSAC symposium on
  Information, computer and communications security}.\hskip 1em plus 0.5em
  minus 0.4em\relax ACM, 2013, pp. 119--130.

\bibitem{laskov2014practical}
P.~Laskov \emph{et~al.}, ``Practical evasion of a learning-based classifier: A
  case study,'' in \emph{Proceedings of the 2014 IEEE Symposium on Security and
  Privacy (S\&P)}.\hskip 1em plus 0.5em minus 0.4em\relax IEEE, 2014, pp.
  197--211.

\bibitem{biggio2013evasion}
B.~Biggio, I.~Corona, D.~Maiorca, B.~Nelson, N.~{\v{S}}rndi{\'c}, P.~Laskov,
  G.~Giacinto, and F.~Roli, ``Evasion attacks against machine learning at test
  time,'' in \emph{Joint European Conference on Machine Learning and Knowledge
  Discovery in Databases}.\hskip 1em plus 0.5em minus 0.4em\relax Springer,
  2013, pp. 387--402.

\bibitem{liang2016cracking}
B.~Liang, M.~Su, W.~You, W.~Shi, and G.~Yang, ``Cracking classifiers for
  evasion: A case study on the google's phishing pages filter,'' in
  \emph{Proceedings of the 25th International Conference on World Wide
  Web}.\hskip 1em plus 0.5em minus 0.4em\relax International World Wide Web
  Conferences Steering Committee, 2016, pp. 345--356.

\bibitem{xu2016automatically}
W.~Xu, Y.~Qi, and D.~Evans, ``Automatically evading classifiers,'' in
  \emph{Proceedings of the 2016 Network and Distributed Systems Symposium},
  2016.

\bibitem{fredrikson2015model}
M.~Fredrikson, S.~Jha, and T.~Ristenpart, ``Model inversion attacks that
  exploit confidence information and basic countermeasures,'' in
  \emph{Proceedings of the 22nd ACM SIGSAC Conference on Computer and
  Communications Security}.\hskip 1em plus 0.5em minus 0.4em\relax ACM, 2015,
  pp. 1322--1333.

\bibitem{wu2016methodology}
X.~Wu, M.~Fredrikson, S.~Jha, and J.~F. Naughton, ``A methodology for
  formalizing model-inversion attacks,'' in \emph{Proceedings of the 2016 IEEE
  Computer Security Foundations Symposium (CSF)}.\hskip 1em plus 0.5em minus
  0.4em\relax IEEE, 2016, pp. 355--370.

\bibitem{bruckner2011stackelberg}
M.~Br{\"u}ckner and T.~Scheffer, ``Stackelberg games for adversarial prediction
  problems,'' in \emph{Proceedings of the 17th ACM SIGKDD international
  conference on Knowledge discovery and data mining}.\hskip 1em plus 0.5em
  minus 0.4em\relax ACM, 2011, pp. 547--555.

\bibitem{bruckner2012static}
M.~Br{\"u}ckner, C.~Kanzow, and T.~Scheffer, ``Static prediction games for
  adversarial learning problems,'' \emph{Journal of Machine Learning Research},
  vol.~13, no. Sep, pp. 2617--2654, 2012.

\bibitem{biggio2011design}
B.~Biggio, G.~Fumera, and F.~Roli, ``Design of robust classifiers for
  adversarial environments,'' in \emph{Proceedings of the 2011 IEEE
  International Conference on Systems, Man, and Cybernetics (SMC)}.\hskip 1em
  plus 0.5em minus 0.4em\relax IEEE, 2011, pp. 977--982.

\bibitem{rodrigues2009robustness}
R.~N. Rodrigues, L.~L. Ling, and V.~Govindaraju, ``Robustness of multimodal
  biometric fusion methods against spoof attacks,'' \emph{Journal of Visual
  Languages \& Computing}, vol.~20, no.~3, pp. 169--179, 2009.

\bibitem{biggio2009multiple}
B.~Biggio, G.~Fumera, and F.~Roli, ``Multiple classifier systems for
  adversarial classification tasks,'' in \emph{Proceedings of the International
  Workshop on Multiple Classifier Systems}.\hskip 1em plus 0.5em minus
  0.4em\relax Springer, 2009, pp. 132--141.

\bibitem{biggio2010multiple}
------, ``Multiple classifier systems for robust classifier design in
  adversarial environments,'' \emph{International Journal of Machine Learning
  and Cybernetics}, vol.~1, no. 1-4, pp. 27--41, 2010.

\bibitem{biggio2010multiple2}
------, ``Multiple classifier systems under attack,'' in \emph{Proceedings of
  the International Workshop on Multiple Classifier Systems}.\hskip 1em plus
  0.5em minus 0.4em\relax Springer, 2010, pp. 74--83.

\bibitem{globerson2006nightmare}
A.~Globerson and S.~Roweis, ``Nightmare at test time: robust learning by
  feature deletion,'' in \emph{Proceedings of the 23rd international conference
  on Machine learning}.\hskip 1em plus 0.5em minus 0.4em\relax ACM, 2006, pp.
  353--360.

\bibitem{kolcz2009feature}
A.~Ko{\l}cz and C.~H. Teo, ``Feature weighting for improved classifier
  robustness,'' in \emph{Proceedings of the 6th Conference on Email and
  Anti-Spam}, 2009.

\bibitem{kantchelian2013approaches}
A.~Kantchelian, S.~Afroz, L.~Huang, A.~C. Islam, B.~Miller, M.~C. Tschantz,
  R.~Greenstadt, A.~D. Joseph, and J.~Tygar, ``Approaches to adversarial
  drift,'' in \emph{Proceedings of the 2013 ACM workshop on Artificial
  intelligence and security}.\hskip 1em plus 0.5em minus 0.4em\relax ACM, 2013,
  pp. 99--110.

\bibitem{liang2017deep}
B.~Liang, H.~Li, M.~Su, P.~Bian, X.~Li, and W.~Shi, ``Deep text classification
  can be fooled,'' \emph{arXiv preprint arXiv:1704.08006}, 2017.

\bibitem{grosse2016adversarial}
K.~Grosse, N.~Papernot, P.~Manoharan, M.~Backes, and P.~McDaniel, ``Adversarial
  examples for malware detection,'' in \emph{European Symposium on Research in
  Computer Security}.\hskip 1em plus 0.5em minus 0.4em\relax Springer, 2017,
  pp. 62--79.

\bibitem{kurakin2016adversarial}
A.~Kurakin, I.~Goodfellow, and S.~Bengio, ``Adversarial examples in the
  physical world,'' \emph{arXiv preprint arXiv:1607.02533}, 2016.

\bibitem{sharif2016accessorize}
M.~Sharif, S.~Bhagavatula, L.~Bauer, and M.~K. Reiter, ``Accessorize to a
  crime: Real and stealthy attacks on state-of-the-art face recognition,'' in
  \emph{Proceedings of the 2016 ACM SIGSAC Conference on Computer and
  Communications Security}.\hskip 1em plus 0.5em minus 0.4em\relax ACM, 2016,
  pp. 1528--1540.

\bibitem{shokri2017Membership}
R.~Shokri, M.~Stronati, C.~Song, and V.~Shmatikov, ``Membership inference
  attacks against machine learning models,'' in \emph{Proceedings of the 2017
  IEEE Symposium on Security and Privacy (S\&P)}.\hskip 1em plus 0.5em minus
  0.4em\relax IEEE, 2017.

\bibitem{Kurakin2017Adversarial}
A.~Kurakin, I.~Goodfellow, and S.~Bengio, ``Adversarial machine learning at
  scale,'' in \emph{Proceedings of the Fifth International Conference on
  Learning Representations}, 2017.

\bibitem{wang2016learning}
Q.~Wang, W.~Guo, K.~Zhang, A.~G. Ororbia~II, X.~Xing, C.~L. Giles, and X.~Liu,
  ``Learning adversary-resistant deep neural networks,'' \emph{arXiv preprint
  arXiv:1612.01401}, 2016.

\bibitem{wang2016Adversary}
Q.~Wang, W.~Guo, K.~Zhang, A.~G. Ororbia~II, X.~Xing, X.~Liu, and C.~L. Giles,
  ``Adversary resistant deep neural networks with an application to malware
  detection,'' in \emph{Proceedings of the 23rd ACM SIGKDD International
  Conference on Knowledge Discovery and Data Mining}.\hskip 1em plus 0.5em
  minus 0.4em\relax ACM, 2017, pp. 1145--1153.

\end{thebibliography}

%
\vspace{-35pt}
\begin{IEEEbiography}[{\includegraphics[width=1in,height=1.25in,clip,keepaspectratio]{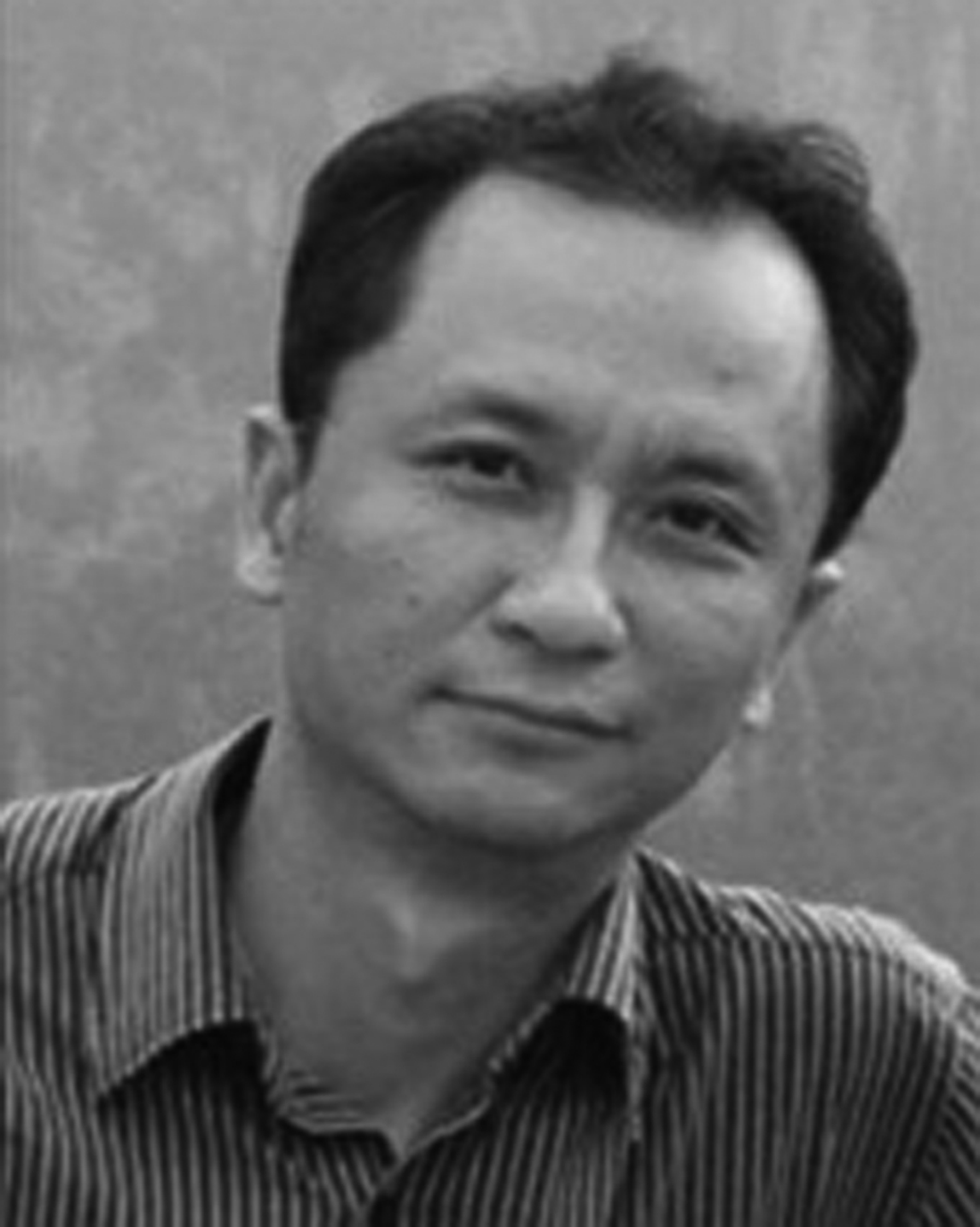}}]{Bin Liang}
received the Ph.D. degree in Computer Science from Institute of Software, Chinese Academy of Sciences. He is currently a professor at School of Information, Renmin University of China. His research interests focus on program analysis, vulnerability detection and Web security.
\end{IEEEbiography}

\vspace{-35pt}
\begin{IEEEbiography}[{\includegraphics[width=1in,height=1.25in,clip,keepaspectratio]{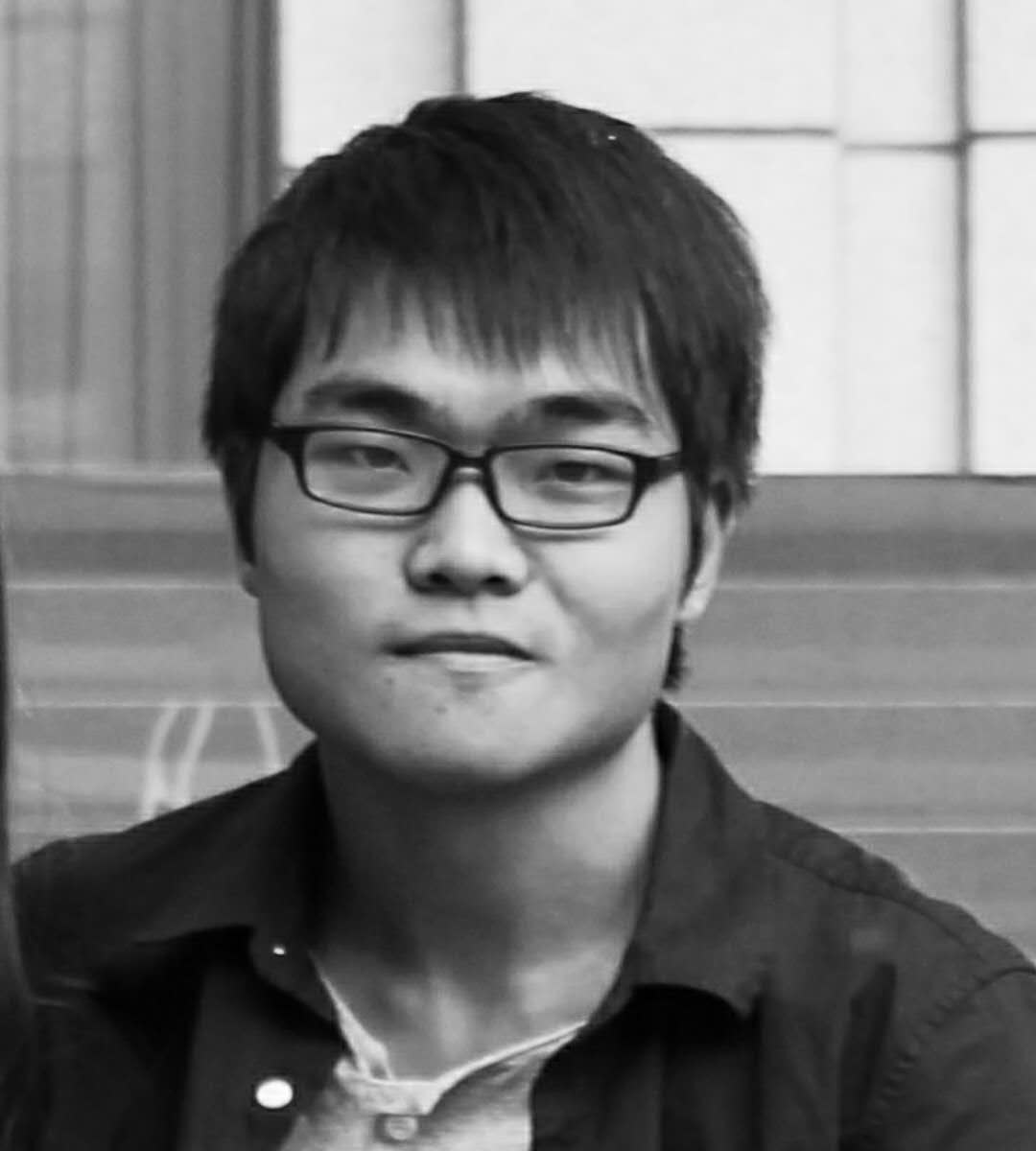}}]{Hongcheng Li}
received the B.S. degree in Information Security from Renmin University of China. He is currently working towards the M.S. degree in Information Security at School of Information, Renmin University of China. His research interests focus on machine learning and security.
\end{IEEEbiography}


\vspace{-35pt}
\begin{IEEEbiography}[{\includegraphics[width=1in,height=1.25in,clip,keepaspectratio]{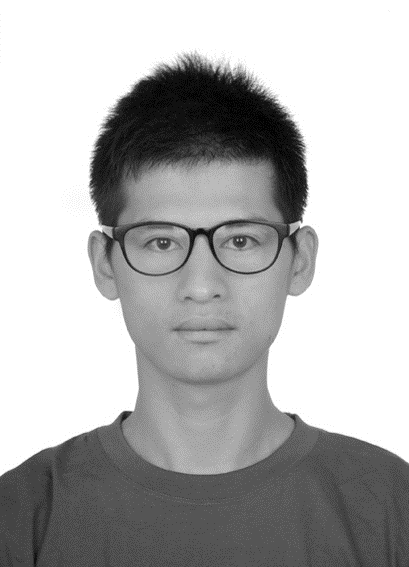}}]{Miaoqiang Su}
received the M.S. degree in Information Security from Renmin University of China. His research interests focus on machine learning.
\end{IEEEbiography}

\vspace{-35pt}
\begin{IEEEbiography}[{\includegraphics[width=1in,height=1.25in,clip,keepaspectratio]{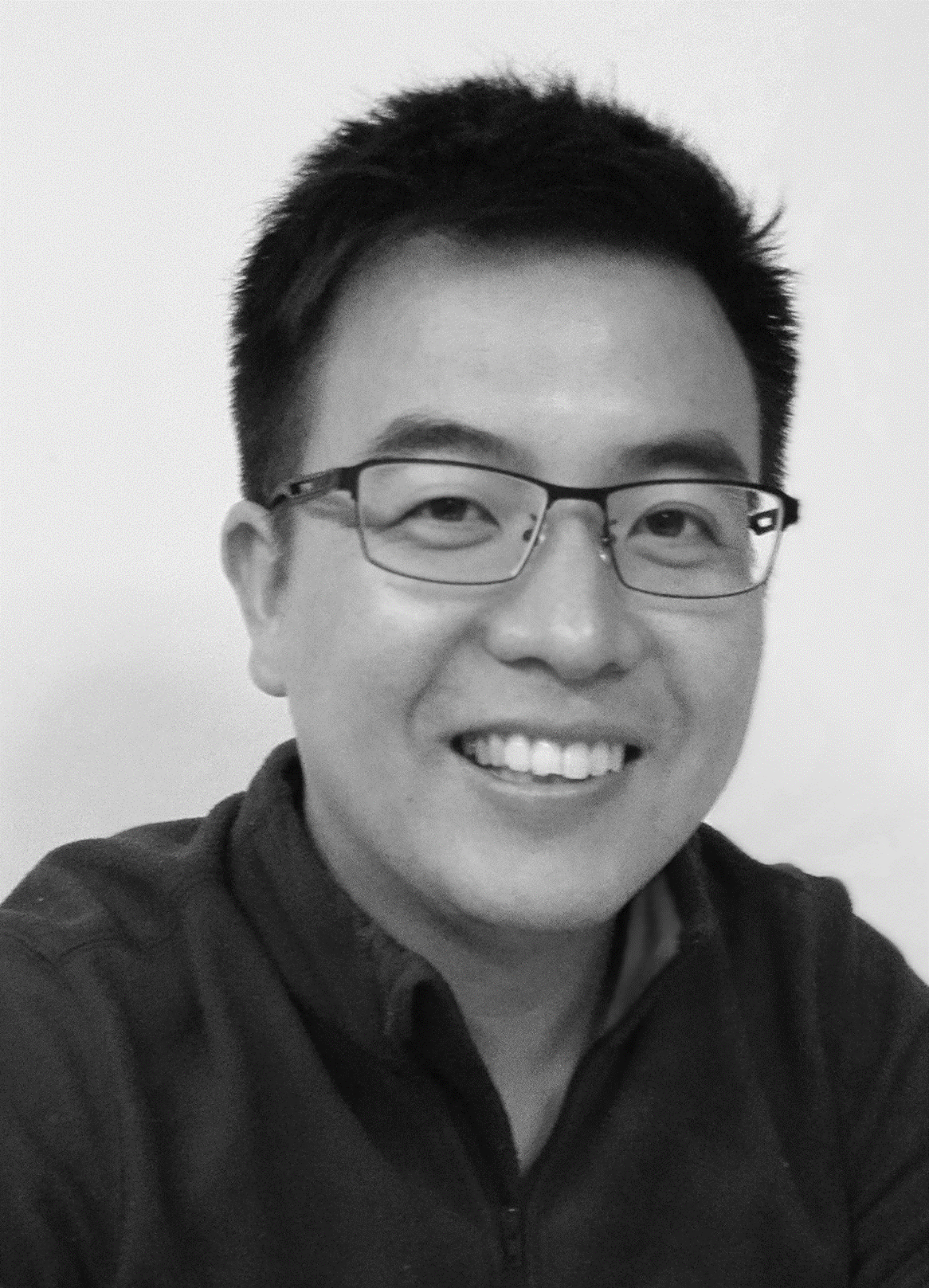}}]{Xirong Li}
received the Ph.D. degree in Computer Science from the University of Amsterdam. He is currently an Associate Professor at School of Information, Renmin University of China. His research interests include multimedia tagging, categorization, and retrieval.
\end{IEEEbiography}

\vspace{-35pt}
\begin{IEEEbiography}[{\includegraphics[width=1in,height=1.25in,clip,keepaspectratio]{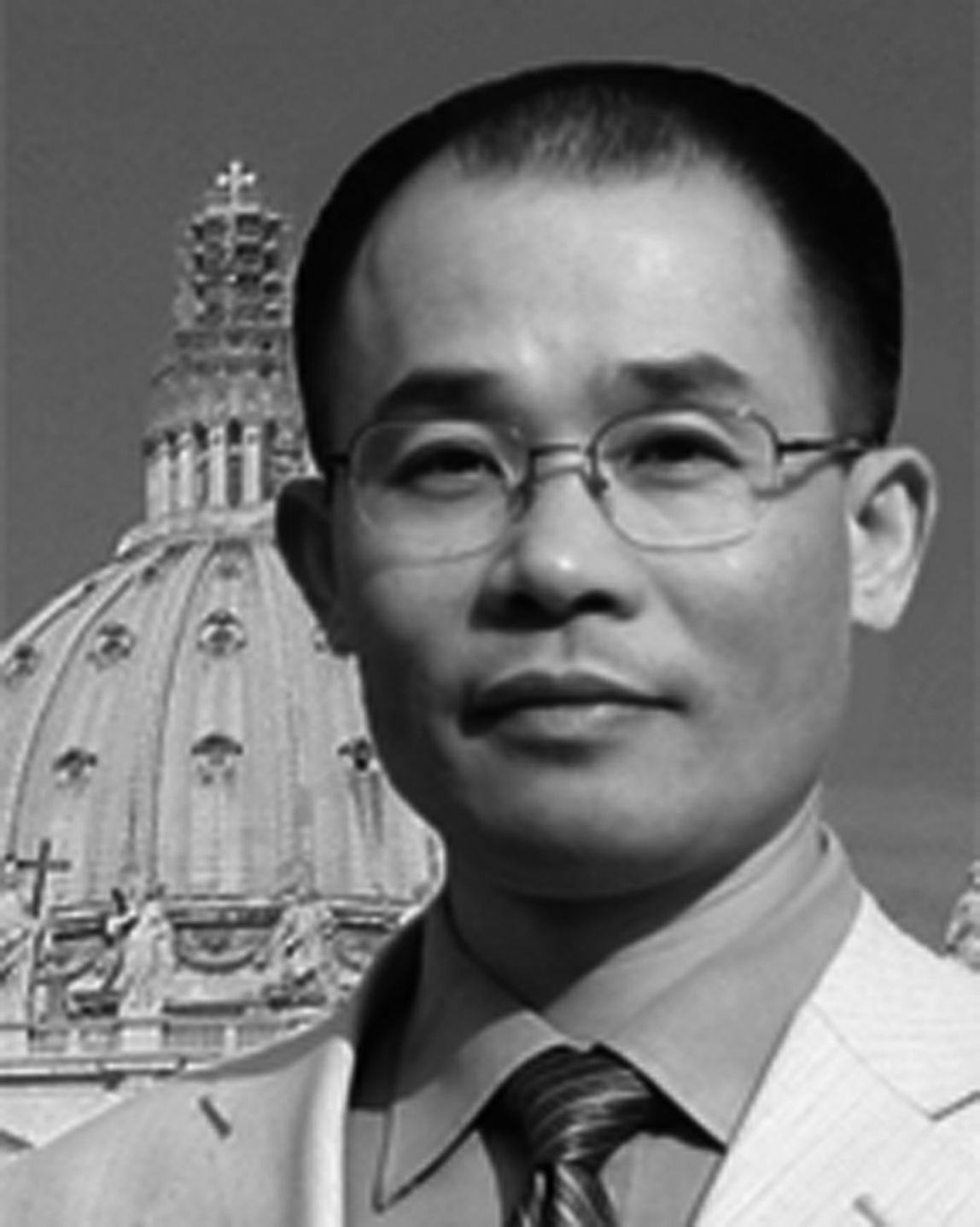}}]{Wenchang Shi}
received the Ph.D. degree in Computer Science from Institute of Software, Chinese Academy of Sciences. He is currently a professor at School of Information, Renmin University of China. His research interests focus on information security, trusted computing, cloud computing, computer forensics, and operating systems.
\end{IEEEbiography}

\vspace{-35pt}
\begin{IEEEbiography}[{\includegraphics[width=1in,height=1.25in,clip,keepaspectratio]{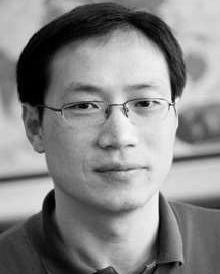}}]{Xiaofeng Wang}
received the Ph.D. degree in computer engineering from Carnegie Mellon University in 2004. Since 2010, he has been the acting Director of the Security Informatics Program at Indiana University Bloomington, where he is currently a Professor with the School of Informatics. His research interests include cloud and mobile security, and data and health informatics security.
\end{IEEEbiography}

\newpage



\end{document}